\documentclass[10pt]{article}
\usepackage{lettrine}

\usepackage[a4paper,margin=2cm]{geometry}
\usepackage{amsmath,amssymb}
\usepackage{graphicx}
\usepackage{subcaption}
\usepackage{booktabs}
\usepackage[colorlinks,citecolor=blue,urlcolor=blue]{hyperref}
\usepackage{authblk}
\usepackage{abstract}
\usepackage{caption}
\usepackage{lineno}
\usepackage[nopatch=footnote]{microtype}
\usepackage{setspace}
\usepackage{adjustbox}

\usepackage{titlesec}
\titleformat{\section}{\normalfont\Large\bfseries\sffamily}{}{0em}{}
\titleformat{\subsection}{\normalfont\large\bfseries\sffamily}{}{0em}{}
\titleformat{\subsubsection}{\normalfont\normalsize\bfseries\fontshape{it}\selectfont}{}{0em}{}
\titlespacing*{\section}{0pt}{1.6em}{0.6em}
\titlespacing*{\subsection}{0pt}{1.2em}{0.4em}
\titlespacing*{\subsubsection}{0pt}{1em}{0.3em}

\newcommand{\sfig}[4]{%
  \begin{figure}[ht]\centering
  \includegraphics[width=#2\linewidth]{#1}
  \caption{#4}\label{#3}\end{figure}}
  
\title{\textsf{\textbf{The power law in human mobility is a mixture artifact: evidence from a pandemic natural experiment}}}

\author[1,2]{Leo Ferres}
\author[2,3]{Bruno Gon\c{c}alves}
\affil[1]{Institute of Data Science, Universidad del Desarrollo, Santiago, Chile}
\affil[2]{ISI Foundation, Turin, Italy}
\affil[3]{Data For Science, Inc, New York, NY}
\date{}

\begin{document}

\maketitle

\begin{abstract}
For nearly two decades, human mobility has been read as scale free.
Displacement distributions follow heavy tails that look like truncated
power laws, traced to individual L\'evy flights. A rival account holds
that movement within each spatial container is lognormal, and the
aggregate power law is an artifact of mixing containers of different
sizes. The two fit the same aggregate data, so the debate has been hard
to settle. We use the COVID-19 lockdowns as a natural experiment that
removes long-distance travel and leaves local travel intact. We
analyze 2.1 billion displacements from 4.4 million mobile-phone users
across three distinct periods. The aggregate exponent increases under lockdown,
from 1.66 to 1.74. Resampling the pre-lockdown traveling population
to match the lockdown population reproduces that shift on its
own, so we must look at the 
individual level to decides the question. 
Lognormal classification is a stable
attractor (81\% retained) while the power-law classification is fragile
(32\% retained), and the users who switch are the ones whose travel
range collapsed the most. Matching the sample size we find that single users' tails reject
the power law and pooled mixtures of equal size pass, so the heavy tail
behavior originates in the aggregate behavior and not on the individual. 
Tail tests find no power-law threshold at full sample size, and the apparent power law disappears
above ten thousand points. A level mixture rebuilds the aggregate
($R^2$ up to 0.98), the radius-of-gyration collapse fails and worsens
under lockdown (CV $= 0.62$ to $0.76$), and the steepening concentrates
in wide-ranging users ($P = 0.0003$). The power law of human travel is a
feature of aggregation, not of individual movement.
\end{abstract}

\section*{Introduction}

How far people travel is a basic quantity in the social and natural sciences\cite{Barbosa_2018}.
It shapes how disease spreads\cite{balcan09-1}, how cities grow\cite{Bettencourt_2007}, and how transport is planned
\cite{Brockmann_2006}. A correct statistical description of travel is needed to
model these processes. The distribution of human displacements is one of the
most replicated findings in computational social science
\cite{Brockmann_2006,Gonzalez_2008,Song_2010}. Over several orders of magnitude
it looks like a power law, $P(\Delta r)\propto \Delta r^{-\alpha}$, with $\alpha$ near
1.6 to 1.8.

Two explanations compete for the origin of this power law. The first holds
that each person moves by a truncated L\'evy flight set by a personal radius
of gyration $r_{g}$ \cite{Gonzalez_2008}. On this view the heavy tail is a real
scale-free property of individual movement. The population power law is that
individual law made visible at scale. We call this H1. The second holds that
movement within any single spatial scale, a neighborhood, a city, a region,
is lognormal, with a length that differs from scale to scale
\cite{Alessandretti_2020}. The population mixes these scales in proportions set
by how far each user ranges. A mixture of lognormals has a lognormal body and
a tail that is hard to distinguish from a power law over a finite sample
\cite{Perline_2005,Bee_2011}. Under this hypothesis, the population power law carries no
information about any individual being scale free. We refer to this as H2.

The two accounts fit aggregate data equally well. Separating them needs an intervention where we
remove one ingredient and keep the other. The COVID-19 lockdowns supplied one by
suppressing long trips between cities and regions while leaving local movement
largely in place\cite{Kraemer_2020}. During this period, the two hypothesis make opposite predictions about the
aggregate exponent. Under H1, $\alpha$ is a property of the individual process
and should stay near constant while under H2, the heavy tail is fed by
long-distance lognormal parts, so removing them should steepen $\alpha$.

We test these predictions with displacement records from Chile across three
lockdown periods. We move past the aggregate exponent, which we show is a weak
test on its own, to individual-level and model-free evidence. The results
converge on H2.

\section*{Results}

\subsection{The pandemic as a container-level ablation}

\begin{figure}[ht]
\centering
\includegraphics[width=.95\textwidth]{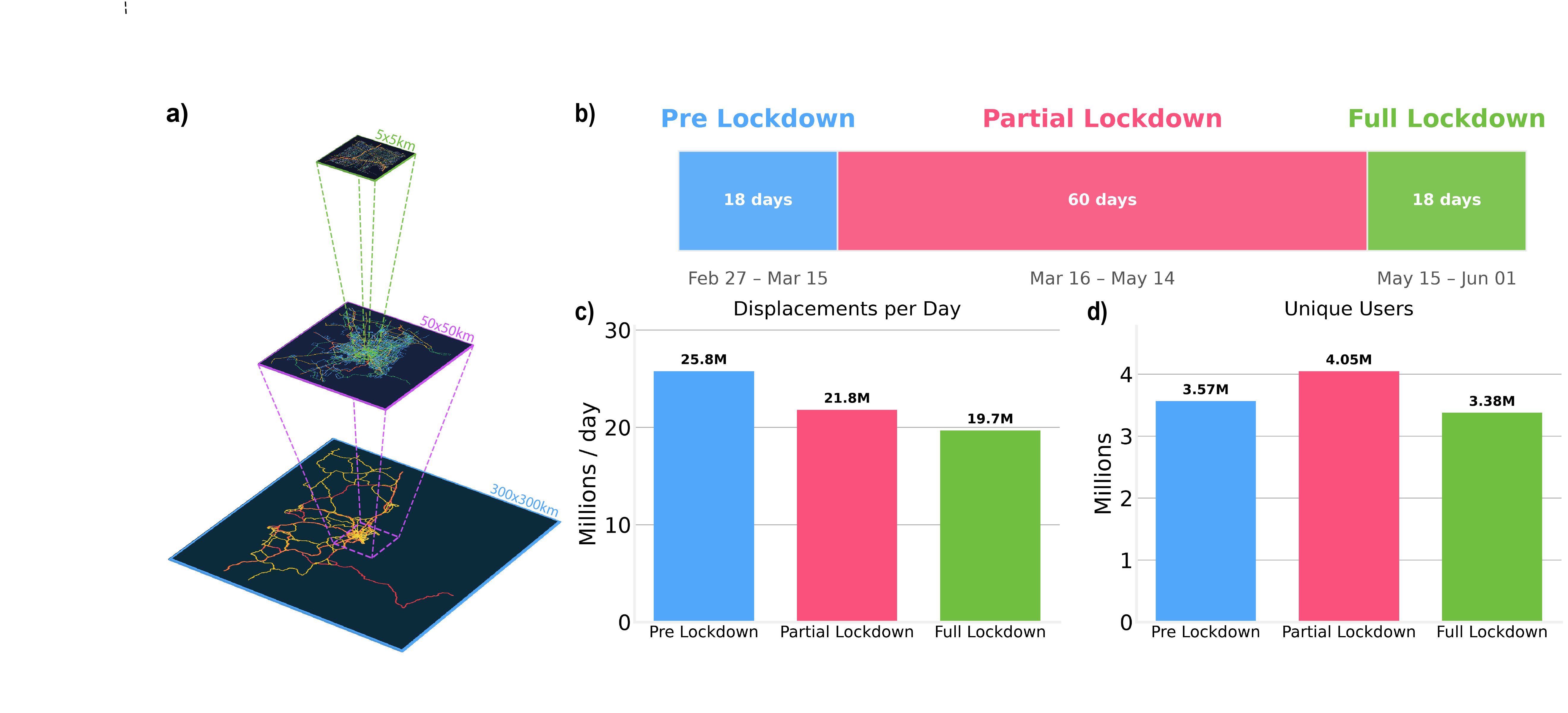}
\caption{\textbf{The dataset} a) We consider three different mobility levels: $5\times 5km$, $50\times 50km$ and $300\times 300km$. During the lockdowns, the largest level is removed. b) The three periods defined by different mobility restrictions. c) Number of displacements per day and d) Number of unique users in each of the three periods.
\label{fig:dataset}}
\end{figure}

\begin{figure*}[ht!]
  \centering
\includegraphics[width=.95\textwidth]{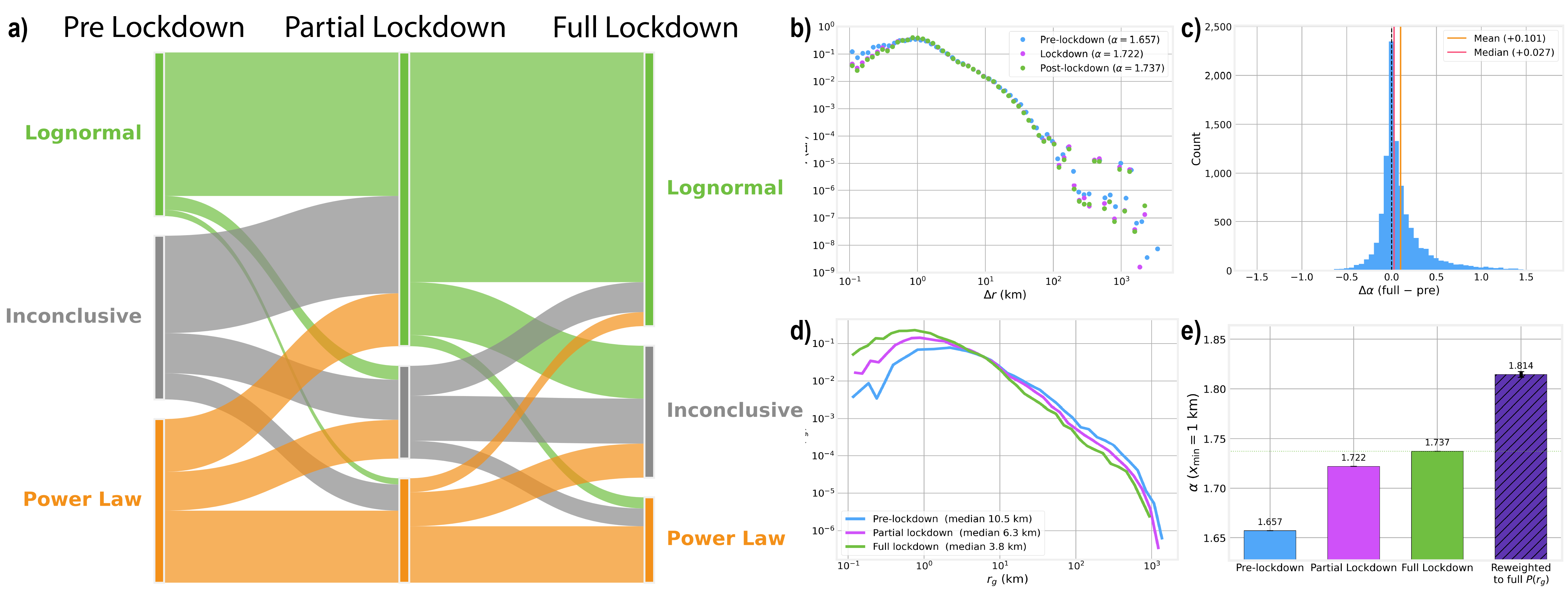}
\caption{\textbf{Impact of the lockdowns} a) As the severity of the lockdowns increases the fraction of users that are best fit by a power law steadily decreases. b) The exponent of the power-law fit increases monotonically. c) Paired individual exponent change ($\Delta\alpha$, full minus
  pre-lockdown).
d) Distribution of the the radius of gyration $r_g$. 
e) Reweighting the pre-lockdown panel to the
  full-lockdown $r_{g}$ distribution, with every displacement still drawn from
   pre-lockdown behaviour, moves $\alpha$ from 1.67 to 1.81}
  
\label{fig:natexp} 
\end{figure*}

We analyze 2.1 billion displacement records from 4.4 million anonymised
mobile-phone users in Chile. The records span three periods (see Fig.~\ref{fig:dataset}) set by government
mobility rules: pre-lockdown (464 million
total displacements), partial lockdown (1.3 billion displacements), 
and full lockdown (354 million displacements). We measure the displacement distance, $\Delta r$, as
the great-circle distance between the cell towers the user was connected to on two consecutive events. The aggregate distribution is well described by a truncated power law in
every period. The exponent $\alpha$ steepens from 1.657 before lockdown to
1.737 under full lockdown, a shift of $+0.080$ (Fig.~\ref{fig:natexp}A). The
median displacement drops from 1.98 to 1.90~km. The median radius of gyration
$r_{g}$ compresses from 7.31 to 2.29~km, a 69\% reduction across 2.86 million
paired users. A likelihood-ratio test at the aggregate level prefers lognormal
over truncated power law in every period ($P\approx 0$). The
direction of the exponent shift matches H2.

The power law exponent is a single number that provides a coarse characterization 
of the mobility distribution, and the change we observe can have different causes. 
To test whether composition alone can produce it, we reweigh pre-lockdown users so that
their $r_{g}$ distribution matches the full-lockdown distribution and pool only
their pre-lockdown displacements before refitting. Despite every displacement still originating
from the pre-lockdown period, the fitted exponent increases from 1.67 to 1.81, well past
the real full-lockdown value of 1.73 (Fig.~\ref{fig:natexp} d)). Changing
who travels reproduces and overshoots the whole steepening, with no change in
individual behavior.\footnote{The composition analysis uses a panel-based fit
whose observed full-minus-pre shift is $+0.065$, and the composition-only
shift is $+0.141\pm0.003$. The headline aggregate shift of $+0.080$ comes from
the population fit. See SI Appendix.} A matched test on simulated H1 worlds
shows that truncated-L\'evy kernels also shift under the same reweighting, by
amounts that bracket the observed value (SI Appendix). The aggregate exponent
is not enough to distinguish between the two hypothesis so we must move to the individual
level.


\subsection{Individual displacements move toward lognormal during lockdown}

We build a panel of 1.52 million users with at least 50 displacements in every
period, fit both a truncated power law and a lognormal to each user-period and
select the better model by a likelihood-ratio test. The fraction of users best described by
a lognormal distribution rises from 49.5\% before lockdown to 60.9\% under full
lockdown while the fraction best described by a truncated power-law falls from 18.1\% to 13.9\%
(Fig.~\ref{fig:natexp}a). The within-user transitions are lopsided. Of users
classed as lognormal before lockdown, 78.7\% stay lognormal under it, and only
4.9\% switch to power law. Of users classed as power law, only 36.9\% keep
that class, and 30.9\% switch to lognormal (Fig.~\ref{fig:natexp}a). The lognormal
class is a stable while the power-law class is fragile. The paired exponent change
has a mean of $+0.102$ (95\% CI $+0.096$ to $+0.107$) and a small effect size
(Cohen's $d=0.349$), but the direction is clear (paired $t=34.8$,
$P=8.8\times10^{-251}$). Both models are simple approximations to a complex phenomenon, 
so these are comparative statements, but the Lognormal model is the more robust fit.


\subsection{The power law belongs originate in the aggregate behavior}

\begin{figure}[ht]
  \centering
  \includegraphics[width=.95\columnwidth]{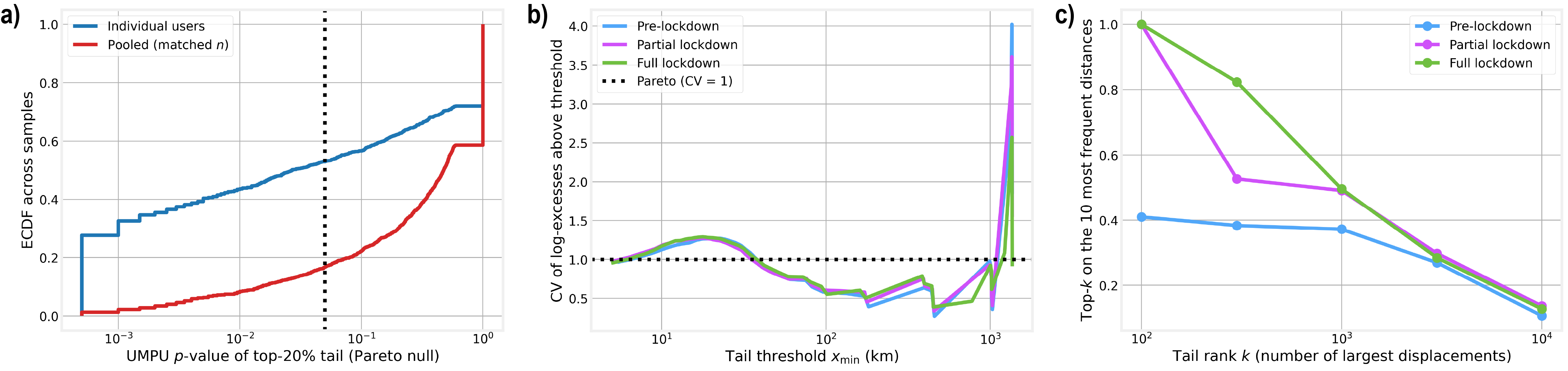}
  \caption{\textbf{Statistical Analysis} a) The distribution of
  maximum-entropy tail-test $p$-values for the top-20\% tail, across individual
  users (blue) and pooled samples of matched size (red). b) The log-excess
  coefficient of variation against the tail threshold, across all three periods. c) Share of the top-$k$ largest displacements carried by the ten most frequent distance values.}
  \label{fig_3}
\end{figure}

A common objection to individual fits is that users have fewer points than the
aggregate. We address this objection by matching sample sizes. For each of 1,537
users with at least 300 pre-lockdown displacements, we compare the user's
top-20\% tail with a pooled population sample of the same size. We test each
tail for a power law with a maximum-entropy method \cite{Bee_2011,del_Castillo_1999} (SI Appendix).
Individual tails reject the power law 52.9\% of the time while pooled samples of the
same size reject it only 16.7\% of the time (Fig.~\ref{fig_3} a). The
pooled power-law tail runs about three times longer (median tail fraction
0.281 against 0.096). At equal sample size, the power-law character originates in
the aggregate across all users. Checks for repeated commute distances and for
tail censoring produce a similar result (SI Appendix).

\subsection{No distance threshold is a power law at full statistical power}

We scan the tail threshold rank by rank at the full two-million sample size,
where the tests have the most power \cite{Bee_2011}. A true power-law tail
would show a log-excess coefficient of variation equal to 1 at every
threshold. The data never settle there (Fig.~\ref{fig_3} b). The curve
is over-dispersed (values above 1) near 10 to 30~km, where city and region lognormals mix, 
and it is under-dispersed (below 1) from 50 to 800~km, inside the container interiors. 
The two isolated points where it crosses 1  sit close to the cutoffs a
fitting library might pick. A two-sided test rejects the power law at 97 to
98\% of thresholds in all three periods. The extreme tail is not a
continuous distribution as the hundred longest displacements in a two-million
sample take only seven distinct values under full lockdown, down from 67
before lockdown (Fig.~\ref{fig_3} c). These are specific city-pair
corridors that collapse onto fewer routes as lockdown constraints between-city
travel. This phenomenon is predicted by H2, while H1 provides no explanation.

\subsection{A container mixture rebuilds the aggregate, and local displacements are lognormal}
\begin{figure*}[ht]
  \centering
  \includegraphics[width=0.55\textwidth]{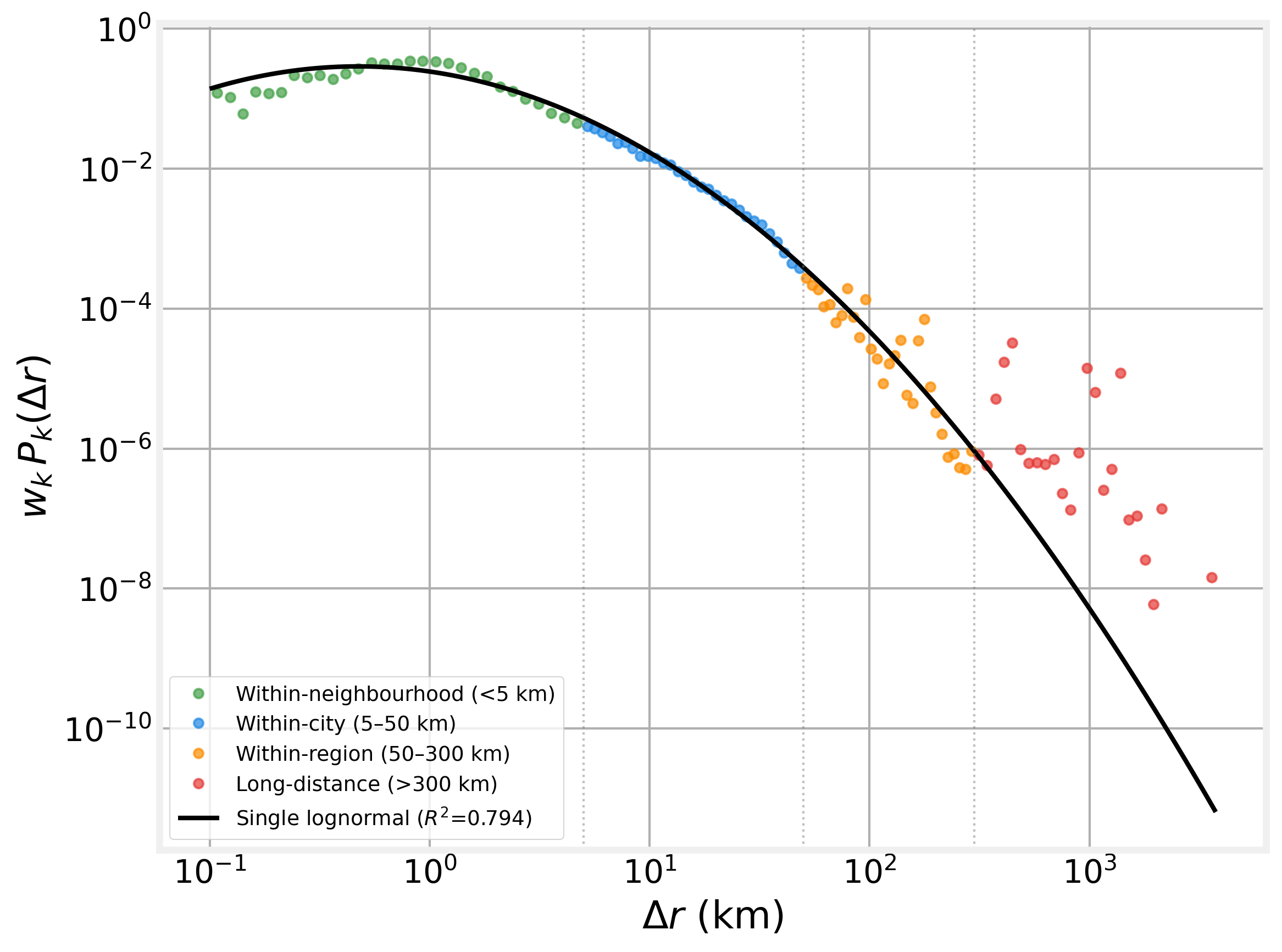}
 \caption{\textbf{Within-level decomposition.}  Within-level displacement distributions (pre-lockdown) with
  lognormal fits.
  }
  \label{fig:decomp}
\end{figure*}

If H2 holds, combining the within-level lognormal densities with the right
weights should rebuild the aggregate. We estimate each level's density by
kernel density estimation in log-space and mix the levels by their observed
weights. The mixture matches the aggregate well (Fig.~\ref{fig:decomp}), with
$R^{2}=0.977$ before lockdown and $R^{2}=0.984$ under full lockdown. We also
fit lognormal and power law within each distance level. At the neighborhood
scale (below 5~km), lognormal wins across every threshold set we explored. At
intermediate scales (5 to 50~km) the picture is mixed, and a fuller analysis
appears in the SI Appendix. These results are futher evidence that local, neighborhood-scale displacements are lognormal and that the scale-free behavior is observable only in the aggregate.

\subsection{The radius-of-gyration does not collapse}
\begin{figure*}[ht]
  \centering
  \includegraphics[width=0.95\textwidth]{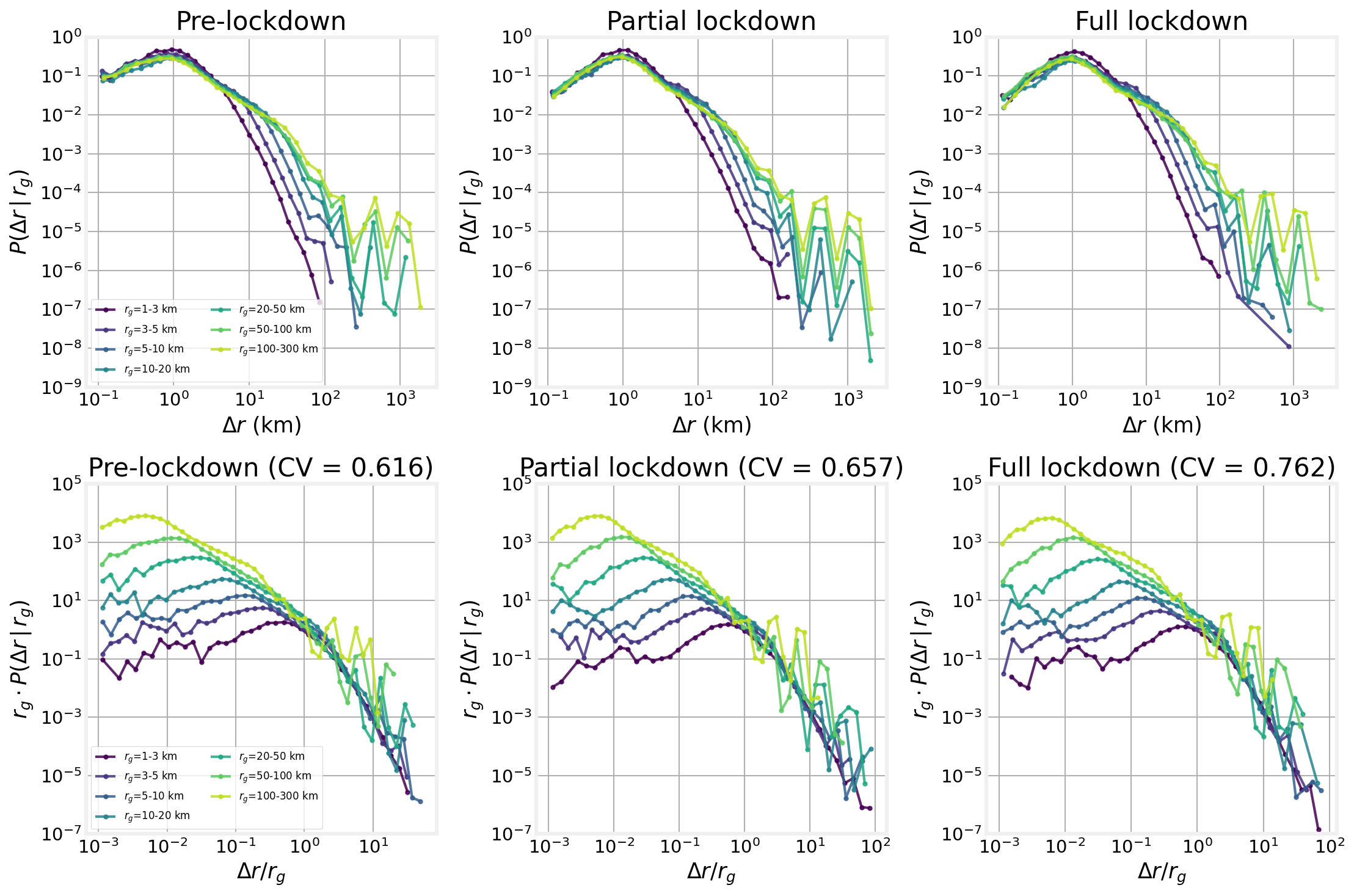}
 \caption{\textbf{Radius of Gyration collapse.} The $r_g$ rescaling collapse across periods. Top
row: conditional distributions by $r_g$ bin. Bottom row: rescaled collapse. The coefficient of variation rises from 0.616 (pre-lockdown) to 0.762 (full lockdown). 
  }
  \label{fig:collapse}
\end{figure*}

The L\'evy flight mechanism makes one sharp quantitative prediction. Rescaling each
user's displacements by $r_{g}$ should collapse them onto a single curve
$F(\Delta r/r_{g})$ \cite{Gonzalez_2008}. We measure collapse quality by the
coefficient of variation across seven $r_{g}$ bins. A good collapse corresponds to a
value well below 0.1. We find the collapse to poor across every period (0.616 before
lockdown) and worsens under partial (0.657) and full lockdown (0.762) (Fig.~\ref{fig:collapse}A). Under
H1 the collapse should not depend on $r_{g}$, while H2  explains the poor collapse through the suppression of the upper container levels. We then split the paired users by
their pre-lockdown $r_{g}$ quartile. The mean steepening rises from $+0.051$ in
the lowest quartile to $+0.184$ in the highest (Fig.~\ref{fig:collapse}B), a
reliable gap (two-sample $t=15.0$, $P<10^{-4}$). High-range users routinely
crossed container boundaries. Lockdown removed those crossings and thinned
their tails. Low-range users were already local and barely moved. The L\'evy
model, in which the exponent belongs to the individual process, gives no
reason for this gradient.

\section*{Discussion}

Using detailed mobility datasets for several million mobile-phone users during the CoVID-19 lockdowns in Chine as a natural experiment, we explore the one of the fundamental results of human mobility.
Our results show that the power-law observed in human mobility is a mixture artifact. 
Individual displacements move toward lognormal during lockdown, and the lognormal class is stable while the power-law class is not. At matched sample size, the power-law tail belongs to
pooled mixtures, not to individuals. At full power, no distance threshold
yields a power-law tail, and the extreme tail is a set of discrete corridors
that thin under lockdown. A container-level mixture rebuilds the aggregate.
The collapse predicted by the L\'evy account fails and worsens during
lockdown. The exponent shift concentrates in users with a large spatial range,
as the container account predicts.

The aggregate exponent steepens, but this shift carries little weight on its
own. Composition alone reproduces the increase of the exponent. The case for H2 rests on
the individual-level and model-free evidence, where it is strong.

Our results help explain why this question remained unsettled for fifteen years.
The apparent power-law tail appears real to any test at the $10^{3}$ to $10^{5}$
sample size of the classic studies, but it disappears at samples sizes of several million (SI Appendix). A direct 
model-comparison still favors lognormal at every sample size. Earlier
studies sat in the range where a lognormal mixture looks like a power law
\cite{Perline_2005}.

These findings extend the container model of Alessandretti and colleagues
\cite{Alessandretti_2020} which showed real trajectories sort into
hierarchical containers with characteristic, non-scale-free sizes. Our
experiment shows these containers matter causally. Removing access to the
upper levels changes the aggregate in the way the container account predicts.
Independent work on the German lockdown found the same selective loss of long
trips \cite{Schlosser_2020} corroborating earlier reports that
within-mode and within-city displacements are not scale free
\cite{Liang_2012,Zhao_2015}.

Trip length within a
container is the product of many positive factors, such as destination type,
routing, and availability. Products of positive factors tend to lognormal\cite{Mitzenmacher_2004}.
This argument is qualitative. Formal normality tests on log displacements
reject exact lognormal behavior (SI Appendix), so the within-level law is close to, while not  exactly lognormal.

Several limits apply. The individual exponent shift is small (Cohen's
$d=0.349$), and about 40\% of users show a decrease, which reflects real
heterogeneity. Our distance thresholds are a proxy for true container
membership, not a direct measure of the spatial hierarchy. The model
classifications depend on the lower cutoff $x_{\min}$, though the lognormal
preference holds across cutoffs (SI Appendix). The data come from one country
during one unusual period, so the exact magnitudes may not transfer. We
analyse displacement sizes without direction or timing. A full-trajectory
container inference combined with the lockdown experiment could sharpen the
test further.

Chile's lockdown was not binary. The Paso a Paso system moved each comuna
through five restriction levels at different times \cite{Pappalardo_2023}. This
staggered roll out supports a dose-response test. The more a comuna's
cross-container travel is suppressed, the more its displacement distribution
should shift toward lognormal. A staggered difference-in-differences design
\cite{Callaway_2021}, paired with a measure of behavioral adherence and a multi-scale robustness protocol, would turn the
binary contrast here into a graded one. We develop this design in future
work.

\section*{Materials and Methods}

\subsection*{Data}

We use anonymised mobile-phone call detail records (CDR) from a major Chilean telecommunications provider\cite{Blondel_2015}. Each record contains a hashed user identifier, a timestamp, and the geographic coordinates of the serving cell tower. We compute displacements as great-circle distances between consecutive tower locations for the same user, excluding zero-distance records (consecutive events at the same tower). The dataset spans February~26 to April~13, 2020, yielding $2.1 \times 10^9$ non-zero displacements from $4.4 \times 10^6$ unique users. We define three study periods based on Chilean government mobility restrictions: pre-lockdown (February~26 -- March~15), partial lockdown (March~16 -- March~25), and full lockdown (March~26 -- April~13). All data were de-identified at the source; the study was approved by the institutional review board of Universidad del Desarrollo.

\subsection*{Radius of gyration}

For each user $i$ observed at tower locations $\{\mathbf{r}_1, \mathbf{r}_2, \ldots, \mathbf{r}_N\}$, the radius of gyration is
\begin{equation}
  r_{g}^{(i)} = \sqrt{\frac{1}{N}\sum_{j=1}^{N} \lVert \mathbf{r}_j - \bar{\mathbf{r}}_i \rVert^2},
\end{equation}
where $\bar{\mathbf{r}}_i = N^{-1}\sum_j \mathbf{r}_j$ is the user's centroid. We compute $r_{g}$ separately for each period. For paired analyses, we restrict to $2.86 \times 10^6$ users present in both the pre-lockdown and full-lockdown periods.

\subsection*{Cross-period panel and sampling}

To ensure that distributional shifts reflect behavioural change rather than compositional turnover, we construct a balanced panel of $1.52 \times 10^6$ users who contribute $\geq 50$ displacements in each of the three periods. From this panel, we draw a primary sample of $10{,}000$ users (random seed~42) for individual-level distributional fitting. A replication sample of $10{,}000$ users (seed~43, $1\%$ overlap) confirms all headline statistics within sampling noise (see Robustness). The minimum threshold of $50$ displacements ensures adequate statistical power for distributional model comparison; convergence analysis (Extended Data) confirms that results stabilise above ${\sim}200$ displacements per user-period.

\subsection*{Distributional fitting}

For each user-period, we fit two candidate models to the empirical displacement distribution above a fixed minimum threshold $x_{min} = 0.5$~km:

\paragraph{Truncated power law (TPL).} The probability density is $p(\Delta r) \propto \Delta r^{-\alpha} \exp(-\Delta r / \lambda)$, where $\alpha > 1$ is the power-law exponent and $\lambda$ is the exponential truncation scale. Parameters are estimated by maximum likelihood.

\paragraph{Lognormal (LN).} The probability density is $p(\Delta r) = (2\pi\sigma^2)^{-1/2}\,\Delta r^{-1}\,\exp\!\bigl[-(\ln \Delta r - \mu)^2 / (2\sigma^2)\bigr]$, with location parameter $\mu$ and scale parameter $\sigma$ estimated by maximum likelihood on displacements above $x_{min}$.

Model comparison uses the Vuong likelihood ratio test for non-nested models. For each user-period, we compute the normalised log-likelihood ratio $R = [\ell_{\mathrm{LN}} - \ell_{\mathrm{TPL}}] / (\sigma_R \sqrt{n})$, where $\ell$ denotes the per-observation log-likelihood and $\sigma_R$ is the standard deviation of the pointwise log-likelihood ratios. We classify the preferred model as lognormal ($R < -z_{0.05}$), truncated power law ($R > z_{0.05}$), or inconclusive ($|R| \leq z_{0.05}$), corresponding to a two-sided significance level of $P < 0.1$.

\paragraph{Sensitivity to $x_{min}$.} The primary analysis uses a fixed $x_{min} = 0.5$~km for all users. We also conduct a per-user $x_{min}$ optimisation via grid search over $\{0.1, 0.2, \ldots, 5.0\}$~km, selecting the value that minimises the Kolmogorov--Smirnov distance between the empirical tail and the fitted model, following the Clauset--Shalizi--Newman framework. The median optimal $x_{min}$ is $1.5$~km, and classification agreement between fixed and data-driven approaches is $43.6\%$. Although the absolute classification fractions change, the qualitative direction of all results --- the shift toward lognormal during lockdown, the transition asymmetry, the $r_{g}$-quartile gradient --- is invariant (see Robustness).

\subsection*{Bootstrap goodness-of-fit}

To assess absolute model adequacy (as opposed to relative model preference), we perform bootstrap Kolmogorov--Smirnov tests. For each user-period, we generate $500$ synthetic datasets from the fitted model, refit, compute the KS statistic for each synthetic dataset, and obtain a $p$-value as the fraction of synthetic KS statistics exceeding the observed value. A model is deemed plausible at the $5\%$ level if $p > 0.05$. Neither lognormal nor truncated power law passes this test for $> 95\%$ of user-periods, confirming that both are approximations. All distributional claims in this paper are therefore framed comparatively.

\subsection*{Distance-level classification}

We classify each displacement into one of four distance levels: neighbourhood ($\Delta r < 5$~km), city ($5 \leq \Delta r < 50$~km), region ($50 \leq \Delta r < 300$~km), and long-distance ($\Delta r \geq 300$~km). These thresholds are motivated by the spatial hierarchy of Chilean urban structure but are not critical to the results: the Robustness section reports that varying thresholds across alternative boundary sets does not alter the qualitative conclusions. At each level, we test lognormal versus power law using the same Vuong likelihood ratio framework applied to the within-level displacement pool.

\subsection*{KDE-based level mixture}

To test whether combining within-level components reproduces the aggregate distribution, we estimate a non-parametric within-level density for each distance level $k$ using Gaussian kernel density estimation in log-space, with bandwidth selected by Silverman's rule\cite{Silverman_1986}. The level-mixture density is
\begin{equation}
  \hat{f}(\Delta r) = \sum_{k=1}^{4} w_k \, \hat{f}_k(\Delta r),
\end{equation}
where $w_k$ is the fraction of total displacements in level $k$ and $\hat{f}_k$ is the KDE estimate for level $k$. We quantify the quality of the mixture reconstruction by $R^2$ between the log-binned aggregate empirical density and the mixture prediction. For comparison, we also fit a free four-component Gaussian mixture model (GMM) in log-space using expectation--maximisation\cite{Dempster_1977}, without constraining component means to correspond to the distance-level boundaries.

\subsection*{Rescaling collapse and collapse quality}

To test H1's prediction of a universal rescaling collapse, we partition users into seven bins by $r_{g}$ (edges at $1, 2, 5, 10, 20, 50, 100, 300$~km) and compute the conditional displacement distribution $P(\Delta r \mid r_{g})$ within each bin. We then form the rescaled distribution $r_{g} \cdot P(\Delta r \mid r_{g})$ as a function of $\Delta r / r_{g}$. Collapse quality is quantified by the coefficient of variation (CV) of the rescaled densities across $r_{g}$ bins, computed at $100$ log-spaced evaluation points in $\Delta r / r_{g} \in [10^{-2}, 10^2]$:
\begin{equation}
  \mathrm{CV} = \frac{1}{M}\sum_{j=1}^{M} \frac{\mathrm{sd}_k[\hat{g}_k(x_j)]}{\mathrm{mean}_k[\hat{g}_k(x_j)]},
\end{equation}
where $\hat{g}_k(x) = r_{g}^{(k)} \cdot P(\Delta r = x \cdot r_{g}^{(k)} \mid r_{g}^{(k)})$ is the rescaled density for $r_{g}$ bin $k$, $M = 100$ is the number of evaluation points, and the standard deviation and mean are taken across the $K = 7$ bins. A perfect collapse yields $\mathrm{CV} = 0$; empirically, $\mathrm{CV} \ll 0.1$ would indicate a satisfactory collapse.

\subsection*{Effect sizes and confidence intervals}

We report Cohen's $d$\cite{Cohen_1988} for paired comparisons as $d = \overline{\Delta\alpha} / s_{\Delta\alpha}$, where $s_{\Delta\alpha}$ is the standard deviation of paired differences. Confidence intervals on population fractions use the Wilson score interval at the $95\%$ level. Bootstrap $95\%$ confidence intervals for the $r_{g}$-quartile gradient are computed from $1{,}000$ bootstrap resamples within each quartile. All reported $P$-values are two-sided unless otherwise noted.

\subsection*{Downsampling placebo}

To rule out the possibility that reduced displacement counts during lockdown bias model classification, we perform a downsampling placebo test. For each user in the $10{,}000$-user panel, we record their lockdown-period displacement count $n_{\mathrm{lock}}$. We then randomly subsample (without replacement) $n_{\mathrm{lock}}$ displacements from the user's pre-lockdown data and refit both distributional models. This procedure is repeated $10$ times, and we report the mean and standard deviation of model-preference fractions across rounds.

\subsection*{Temporal placebo}

To verify that the pre-lockdown period provides a stable baseline, we split it at the midpoint (March~7) and perform the full individual-level distributional analysis on each half independently. We report model-preference fractions, the Pearson correlation of individual $\alpha$ values across halves, and the paired $t$-test for $\Delta\alpha$ between halves.

\subsection*{Spatial robustness (MAUP)}

We assess sensitivity to the modifiable areal unit problem\cite{Fotheringham_1991} by coarsening the spatial resolution of displacements in two ways: (i)~rounding displacement distances to the nearest $\delta \in \{0.5, 1.0, 2.0, 5.0\}$~km, and (ii)~remapping cell-tower coordinates to grid centroids at resolutions of $\{1, 2, 5, 10\}$~km before recomputing displacements. We then repeat the aggregate and individual-level model comparisons at each resolution.

\subsection*{Computational infrastructure}

All analyses were conducted in Python~3.9 using \texttt{scipy}\cite{Virtanen_2020}, \texttt{numpy}\cite{Harris_2020}, \texttt{pandas}\cite{McKinney_2010}, and \texttt{powerlaw} (v1.5)\cite{Alstott_2014}. Kernel density estimation used \texttt{scipy.stats.gaussian\_kde}. Gaussian mixture models were fit using \texttt{scikit-learn}\cite{Pedregosa_2011}. Individual-level fitting was parallelised across $48$ CPU cores; the full analysis pipeline (all periods, all robustness checks, both $10{,}000$-user samples) completes in approximately $12$ hours.


\bibliographystyle{abbrv}
\bibliography{references}

\clearpage
\onecolumn
\newgeometry{margin=1in}
\setcounter{section}{0}
\setcounter{figure}{0}
\setcounter{table}{0}
\renewcommand{\thesection}{S\arabic{section}}
\renewcommand{\thefigure}{S\arabic{figure}}
\renewcommand{\thetable}{S\arabic{table}}

\begin{center}
{\Large\bfseries Supporting Information}\\[5pt]
{\large The aggregate power law in human mobility is a mixture artifact:
evidence from a pandemic natural experiment}\\[5pt]
{\normalsize Leo Ferres \& Bruno Gon\c{c}alves}
\end{center}
\vspace{0.8em}

\noindent
This document holds the data description, the fitting and tail-test methods,
and the robustness checks named in the main text. Figures and tables carry the
detail. Sections S1 to S7 cover data and methods. Sections S8 to S12 test the
individual-level analysis. Sections S13 to S15 test the aggregate analysis.
Sections S16 to S24 cover the mechanism and the model-free tail tests.

\section{Data and periods}
We analyse anonymised mobile-phone displacement records from Chile. The
records cover 27 February to 1 June 2020. A displacement is the great-circle
distance between the serving towers of two consecutive events for one user. We
drop zero-distance pairs and gaps over 24 hours. Three policy periods define
the data (Table~\ref{tab:periods}).

\begin{table}[ht]\centering
\caption{The three lockdown periods and their record and user counts.}
\label{tab:periods}
\begin{adjustbox}{width=\linewidth}
\begin{tabular}{llrr}
\toprule
Period & Dates & Displacements & Users \\
\midrule
Pre-lockdown    & 27 Feb -- 15 Mar 2020 & 463{,}960{,}240   & 3{,}567{,}734 \\
Partial lockdown& 16 Mar -- 14 May 2020 & 1{,}308{,}564{,}121 & 4{,}048{,}417 \\
Full lockdown   & 15 May -- 1 Jun 2020  & 354{,}268{,}789   & 3{,}378{,}757 \\
\bottomrule
\end{tabular}
\end{adjustbox}
\end{table}

\section{Sampling strategy}
The full dataset holds 2.1 billion records, too many to load into Python at
once. For aggregate fits we draw 2 million displacements per period with Spark
fraction sampling (seed 42). For the displacement-$r_{g}$ joint analysis we draw
2 million joined records per period. We cache all samples and reuse them across
experiments. For individual fits we build a panel of 1{,}520{,}570 users with
at least 50 displacements in every period.

\section{Distribution fitting}
We fit continuous power-law, truncated power-law, and lognormal models with the
powerlaw library \cite{Alstott_2014,Clauset_2009}. Aggregate fits use
$x_{\min}=1$~km. Individual fits use $x_{\min}=0.5$~km. Model choice uses the
Vuong likelihood-ratio test for non-nested models \cite{Vuong_1989}. A positive
statistic favours the first model, and $P<0.1$ marks a decisive result. We
label a user-period by the preferred model, or inconclusive when the test is
not decisive. Goodness of fit in log-space uses
\[
R^{2} = 1 - \frac{\sum_i (\log_{10} h_i - \log_{10}\hat{f}(x_i))^2}
{\sum_i (\log_{10} h_i - \overline{\log_{10} h})^2},
\]
where $h_i$ are histogram densities at bin centres $x_i$ and $\hat f$ is the
model density.

\section{Distance levels and the level mixture}
We sort displacements into four levels by size: neighbourhood (below 5~km),
city (5 to 50~km), region (50 to 300~km), and long-distance (above 300~km).
These thresholds differ from the Alessandretti boundaries of [3, 27, 88, 162]
km \cite{Alessandretti_2020}. The Santiago tail is sparser, so wider bins keep
each level well populated. For each level we estimate the density by Gaussian
kernel density estimation on log displacements. The mixture density is
\[
f(\Delta r) = \sum_l w_l\, \hat f_l(\Delta r),
\]
where $w_l$ is the fraction of displacements at level $l$ and $\hat f_l$ is the
level density, changed from log-space by $f_{\Delta r}(x)=f_{\log \Delta r}(\log x)/x$.

\section{Collapse quality}
We bin users by $r_{g}$ at edges 1, 3, 5, 10, 20, 50, 100, and 300~km. For each
bin we form the rescaled density $r_{g}\,P(\Delta r\mid r_{g})$ against $\Delta r/r_{g}$. We
interpolate onto a common log-spaced grid of 40 points. We report the mean
coefficient of variation across grid points with at least three contributing
bins. A lower value marks a better collapse.

\section{Effect sizes and intervals}
Cohen's $d$ for the paired exponent shift is the mean of individual differences
over their standard deviation. We compute bootstrap 95\% intervals with
10{,}000 resamples (seed 42) on the mean and median of $\Delta\alpha$ and on
retention rates. We use Wilson score intervals for proportions
\cite{Wilson_1927}.

\section{Computational infrastructure}
Data extraction, joins, and sampling run on Apache Spark 4.1\cite{Zaharia_2016} in local mode,
with 38 cores and 240~GB driver memory, on one on-premise server. Fitting and
plots use Python 3.10 with NumPy, SciPy, scikit-learn, powerlaw, and
matplotlib\cite{Hunter_2007}.

\section{Goodness of fit}
The likelihood-ratio test names the better of two models. It does not say
either model fits well. We refit 2{,}000 users and run absolute tests:
Kolmogorov--Smirnov against the fitted lognormal and power-law distributions,
and Anderson--Darling\cite{Anderson_1954} on log displacements. Among users with a decisive winner,
only 4.7\% pass the KS test for their winning model before and during full
lockdown, and 0.7\% during partial lockdown (Fig.~\ref{fig:ks}). Neither model
is a full description of the individual data. The comparative results still
hold. Lognormal is the better approximation, not a complete one.

\sfig{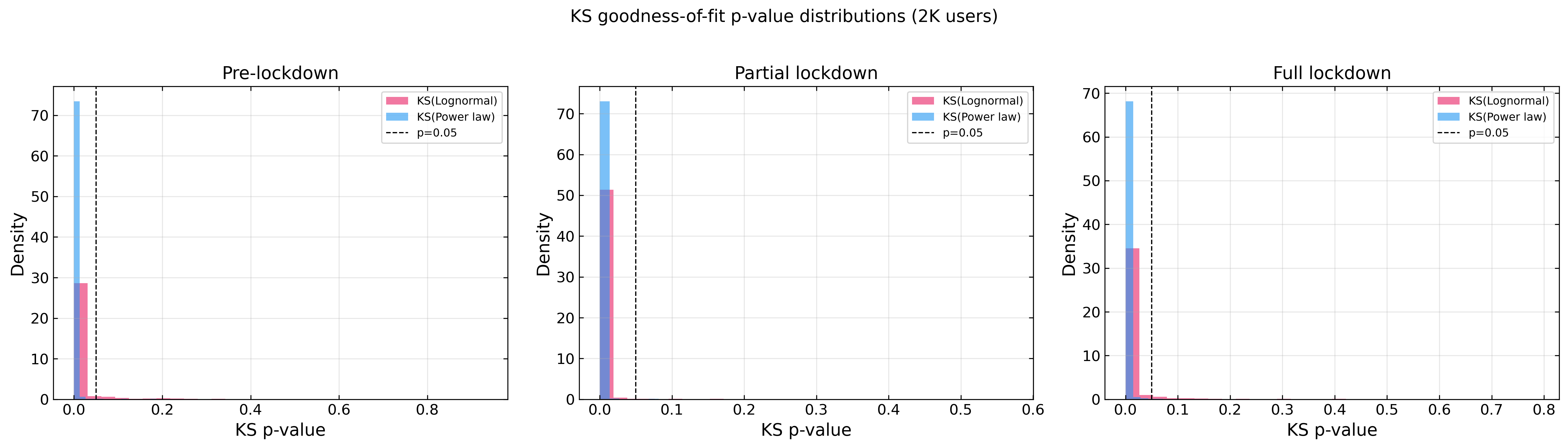}{0.95}{fig:ks}{KS $p$-value distributions for the lognormal (pink) and power-law (blue) fits, by period. The dashed line marks $p=0.05$. Almost all $p$-values fall below 0.05 for both models.}

\section{Sensitivity to the lower cutoff}
Power-law fits depend on the lower cutoff $x_{\min}$. We test
$x_{\min}\in\{0.25,0.5,1.0\}$~km for 200 users. Lognormal beats power law at
every cutoff (Table~\ref{tab:xmin}). We also test a data-driven cutoff per user
by grid search over ten values, picking the one that minimises the KS distance.
The median data-driven cutoff is 1.50~km, above the fixed 0.5~km
(Fig.~\ref{fig:xmin}). Classification agreement between the fixed and
data-driven cutoff is 43.6\% (Cram\'er's $V=0.307$\cite{Cramer_1946}). The direction of all
results holds across cutoffs, but the exact fractions depend on the cutoff. As
the minimum displacement count per user rises from 30 to 300, the lognormal
fraction rises from 49.5\% to 59.8\%, and the power-law fraction stays near 16
to 18\% (Table~\ref{tab:threshold}).

\begin{table}[ht]\centering
\caption{Model fractions across the lower cutoff $x_{\min}$ (pre-lockdown,
$n=200$). Lognormal leads at every cutoff.}
\label{tab:xmin}
\begin{adjustbox}{width=\linewidth}
\begin{tabular}{rrrrr}
\toprule
$x_{\min}$ (km) & TPL \% & LN \% & Inconclusive \% & Median $\alpha$ \\
\midrule
0.25 & 10.5 & 72.0 & 17.5 & 1.444 \\
0.50 & 17.0 & 52.0 & 31.0 & 1.612 \\
1.00 & 21.0 & 30.5 & 48.5 & 1.825 \\
\bottomrule
\end{tabular}
\end{adjustbox}
\end{table}

\begin{table}[ht]\centering
\caption{Model fractions across the minimum displacement count per user
(pre-lockdown, 10{,}000 fits). More data per user strengthens lognormal.}
\label{tab:threshold}
\begin{adjustbox}{width=\linewidth}
\begin{tabular}{rrrrr}
\toprule
Min count & $n$ users & TPL \% & LN \% & Inc.\ \% \\
\midrule
30  & 9{,}995 & 18.1 & 49.5 & 32.4 \\
50  & 9{,}995 & 18.1 & 49.5 & 32.4 \\
100 & 8{,}329 & 18.1 & 51.2 & 30.6 \\
200 & 4{,}112 & 17.3 & 55.5 & 27.2 \\
300 & 1{,}537 & 16.1 & 59.8 & 24.1 \\
\bottomrule
\end{tabular}
\end{adjustbox}
\end{table}

\sfig{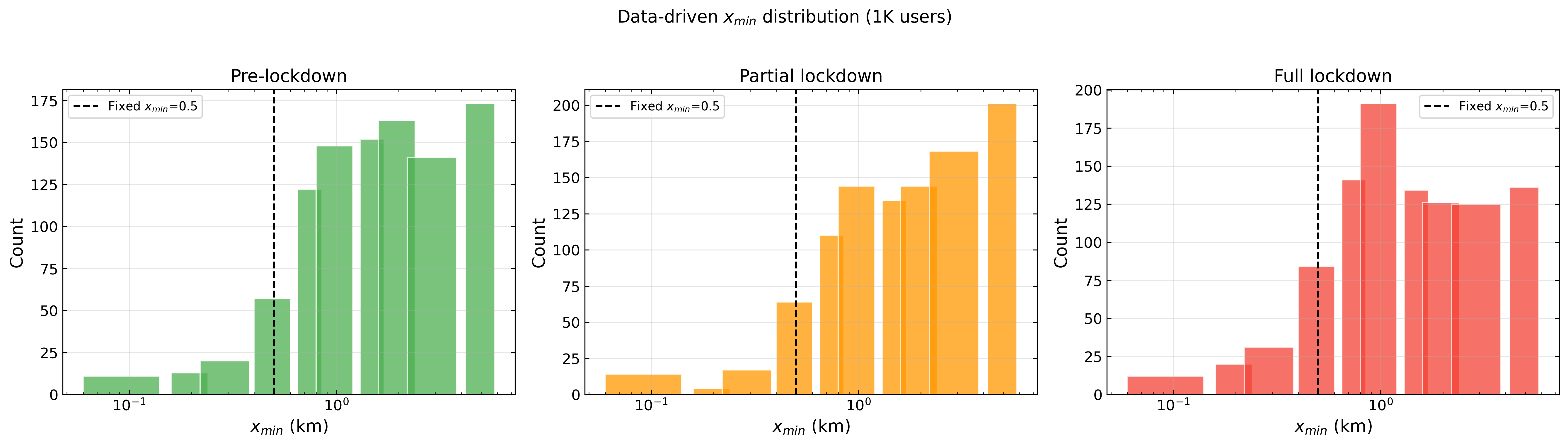}{0.95}{fig:xmin}{Distribution of the per-user data-driven $x_{\min}$ by period. The dashed line marks the fixed $x_{\min}=0.5$~km used in the main text.}

\section{Replication on 10,000 users}
We re-ran the full individual analysis on a 10{,}000-user sample drawn with a
fresh seed (seed 43, 1\% overlap with the original). All headline numbers
reproduce within sampling noise (Table~\ref{tab:rep}). The lognormal fraction
is 49.5\% before lockdown and 60.9\% under full lockdown. The retention
asymmetry holds, with lognormal retention 78.7\% and power-law retention
36.9\%. The mean exponent shift is $+0.101$ with Cohen's $d=0.349$. The $r_{g}$
quartile gradient holds with $P<10^{-4}$. The 10{,}000-user sample is the one
reported in the main text (Fig.~\ref{fig:rep}).

\begin{table}[ht]\centering
\caption{Headline statistics for the original 991-user sample and the
independent 10{,}000-user sample.}
\label{tab:rep}
\begin{adjustbox}{width=\linewidth}
\begin{tabular}{lrr}
\toprule
Statistic & $n=991$ & $n=10{,}000$ \\
\midrule
LN fraction (pre-lockdown)    & 48.7\% & 49.5\% \\
LN fraction (full lockdown)   & 61.4\% & 60.9\% \\
TPL fraction (pre-lockdown)   & 19.8\% & 18.1\% \\
LN$\to$LN retention           & 80.9\% & 78.7\% \\
TPL$\to$TPL retention         & 32.3\% & 36.9\% \\
Mean $\Delta\alpha$           & $+0.103$ & $+0.101$ \\
Median $\Delta\alpha$         & $+0.025$ & $+0.027$ \\
Cohen's $d$                   & 0.358 & 0.349 \\
Q4 vs Q1 $\Delta\alpha$ $P$   & 0.0003 & $<10^{-4}$ \\
\bottomrule
\end{tabular}
\end{adjustbox}
\end{table}

\sfig{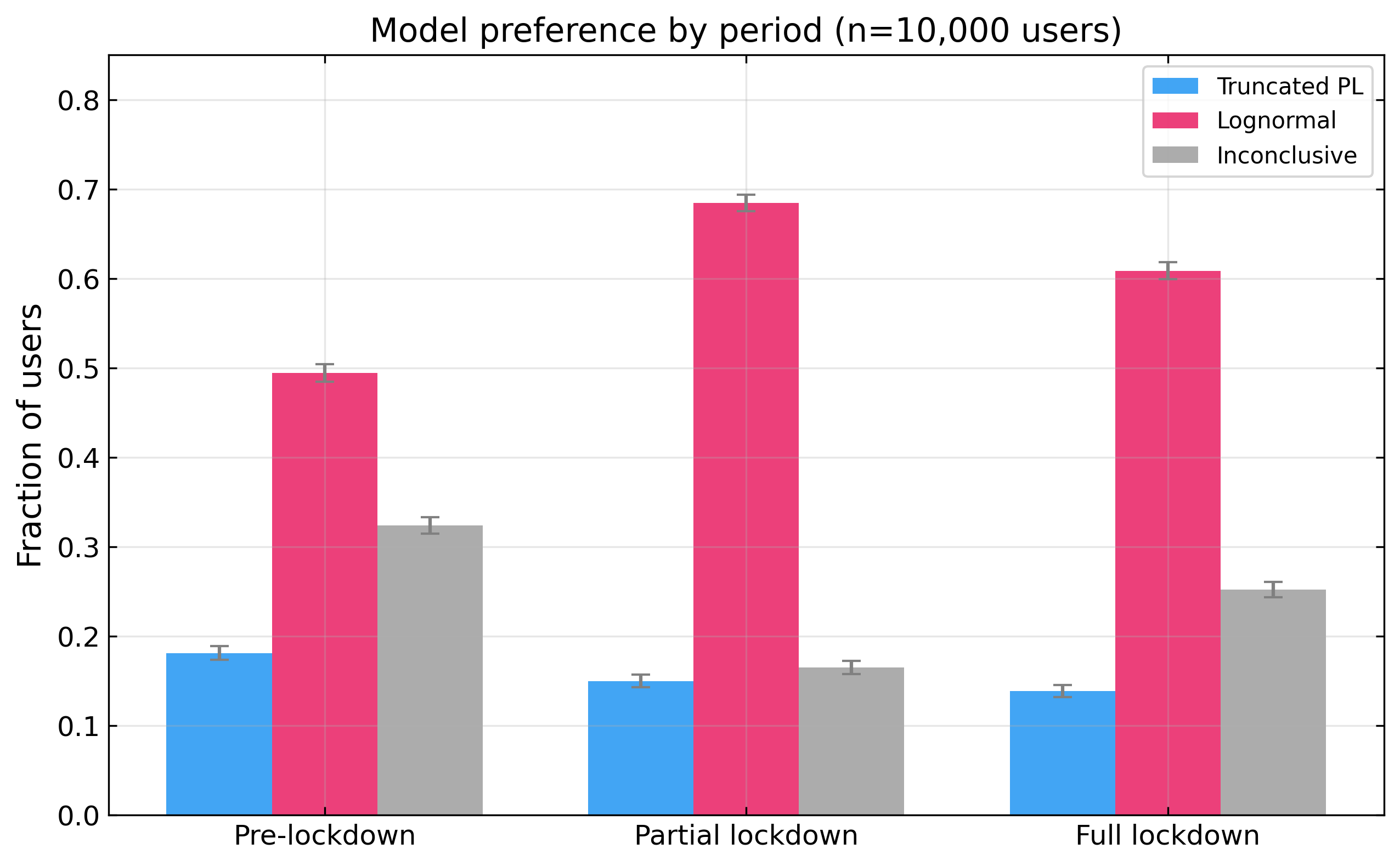}{0.5}{fig:rep}{Winner fractions by period for the 10{,}000-user
sample. Error bars show Wilson 95\% intervals.}

\section{Downsampling test}
A reader could worry that the lockdown shift is an effect of data volume. We
test this directly. For each of 10{,}000 users present in both periods, we
subsample the pre-lockdown displacements to match the full-lockdown count, then
refit. We repeat this ten times. The downsampled pre-lockdown fractions stay
close to the original pre-lockdown values and far from the full-lockdown values
(Table~\ref{tab:ds}, Fig.~\ref{fig:ds}). The lognormal fraction drops from
49.5\% to 45.5\%, an order of magnitude smaller than the lockdown shift. The
median exponent moves from 1.720 to 1.700, against 1.807 under full lockdown.
The shift toward lognormal is not a data-volume effect.

\begin{table}[ht]\centering
\caption{Downsampling test. Winner fractions for the original pre-lockdown
data, the downsampled pre-lockdown data (mean $\pm$ std over ten rounds), and
full lockdown, on the same user set.}
\label{tab:ds}
\begin{adjustbox}{width=\linewidth}
\begin{tabular}{lrrr}
\toprule
Model & Original pre & Downsampled pre & Full lockdown \\
\midrule
Truncated PL & 18.1\% & $17.1\pm0.2$\% & 13.9\% \\
Lognormal    & 49.5\% & $45.5\pm0.2$\% & 60.9\% \\
Inconclusive & 32.4\% & $37.4\pm0.3$\% & 25.2\% \\
\bottomrule
\end{tabular}
\end{adjustbox}
\end{table}

\sfig{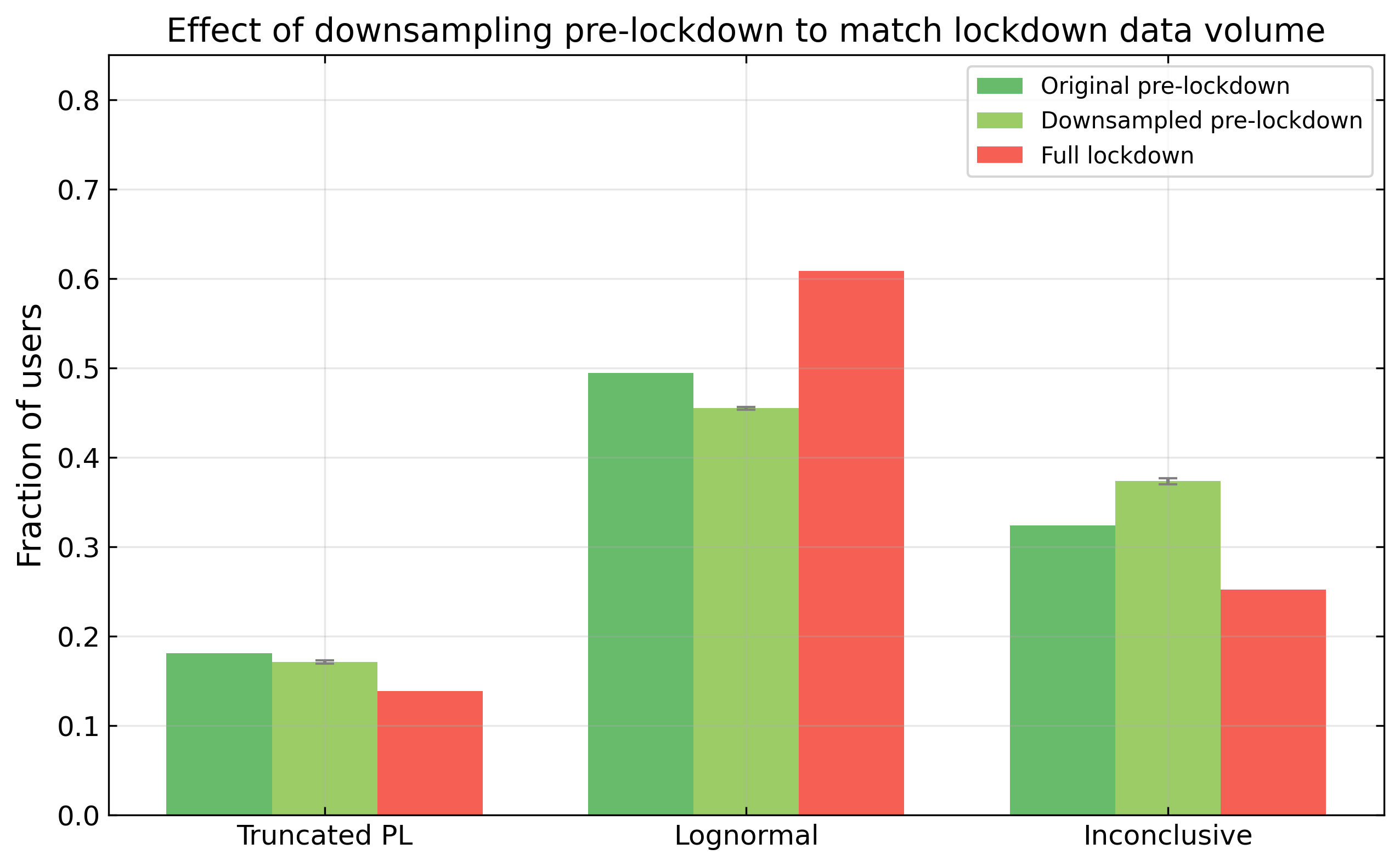}{0.5}{fig:ds}{Downsampling comparison. The downsampled
pre-lockdown curve (light green) stays close to the original (dark green) and
far from full lockdown (red).}

\section{Temporal splitting}
We split the pre-lockdown period at its midpoint and fit each half on its own,
for 8{,}854 users with at least 30 displacements in both halves. The winner
fractions match across halves, with power law 17.3\% against 17.5\%. Individual
exponents correlate well between halves (Pearson $r=0.828$), with no reliable
paired difference (mean shift $+0.002$, $P=0.39$). The pre-lockdown baseline is
stable in time.

\section{Spatial aggregation (MAUP)}
Cell-phone data are discretised by towers, so spatial aggregation could shape
the tail. We test this two ways. First, we round displacement distances to 0.5,
1.0, 2.0, and 5.0~km and refit at the aggregate level. Lognormal wins at almost
every resolution across all three periods (Table~\ref{tab:round}), with one
exception at 1~km. Second, we map the 4{,}858 towers to grid cells of 1, 2, 5,
and 10~km and recompute distances between cell centroids. Lognormal wins at
every grid resolution (Table~\ref{tab:grid}). At the individual level, coarser
resolution strengthens the lognormal preference (Fig.~\ref{fig:indivres}). The
aggregate result holds across spatial units.

\begin{table}[ht]\centering
\caption{Aggregate fits at different rounding resolutions (pre-lockdown).
LN marks a lognormal win ($R<0$, $P<0.1$).}
\label{tab:round}
\begin{adjustbox}{width=\linewidth}
\begin{tabular}{lrrrl}
\toprule
Resolution & $n$ remaining & $\alpha$ & $R$(TPL/LN) & Winner \\
\midrule
Original & 2{,}029{,}724 & 1.752 & $-8{,}174$   & LN \\
0.5 km   & 1{,}983{,}044 & 1.846 & $-530$       & LN \\
1.0 km   & 1{,}869{,}394 & 1.940 & $+159$       & TPL \\
2.0 km   & 1{,}528{,}032 & 1.700 & $-38{,}036$  & LN \\
5.0 km   & 858{,}534     & 1.467 & $-186{,}354$ & LN \\
\bottomrule
\end{tabular}
\end{adjustbox}
\end{table}

\begin{table}[ht]\centering
\caption{Aggregate fits using tower grid-cell remapping (pre-lockdown).}
\label{tab:grid}
\begin{adjustbox}{width=\linewidth}
\begin{tabular}{lrrrl}
\toprule
Grid & $n$ displacements & $\alpha$ & $R$(TPL/LN) & Winner \\
\midrule
Original & 500{,}000     & 1.738 & $-7{,}777$   & LN \\
1 km     & 9{,}103{,}295 & 1.844 & $-235$       & LN \\
2 km     & 7{,}748{,}704 & 1.673 & $-44{,}305$  & LN \\
5 km     & 5{,}423{,}518 & 1.475 & $-198{,}033$ & LN \\
10 km    & 3{,}721{,}244 & 1.378 & $-273{,}618$ & LN \\
\bottomrule
\end{tabular}
\end{adjustbox}
\end{table}

\sfig{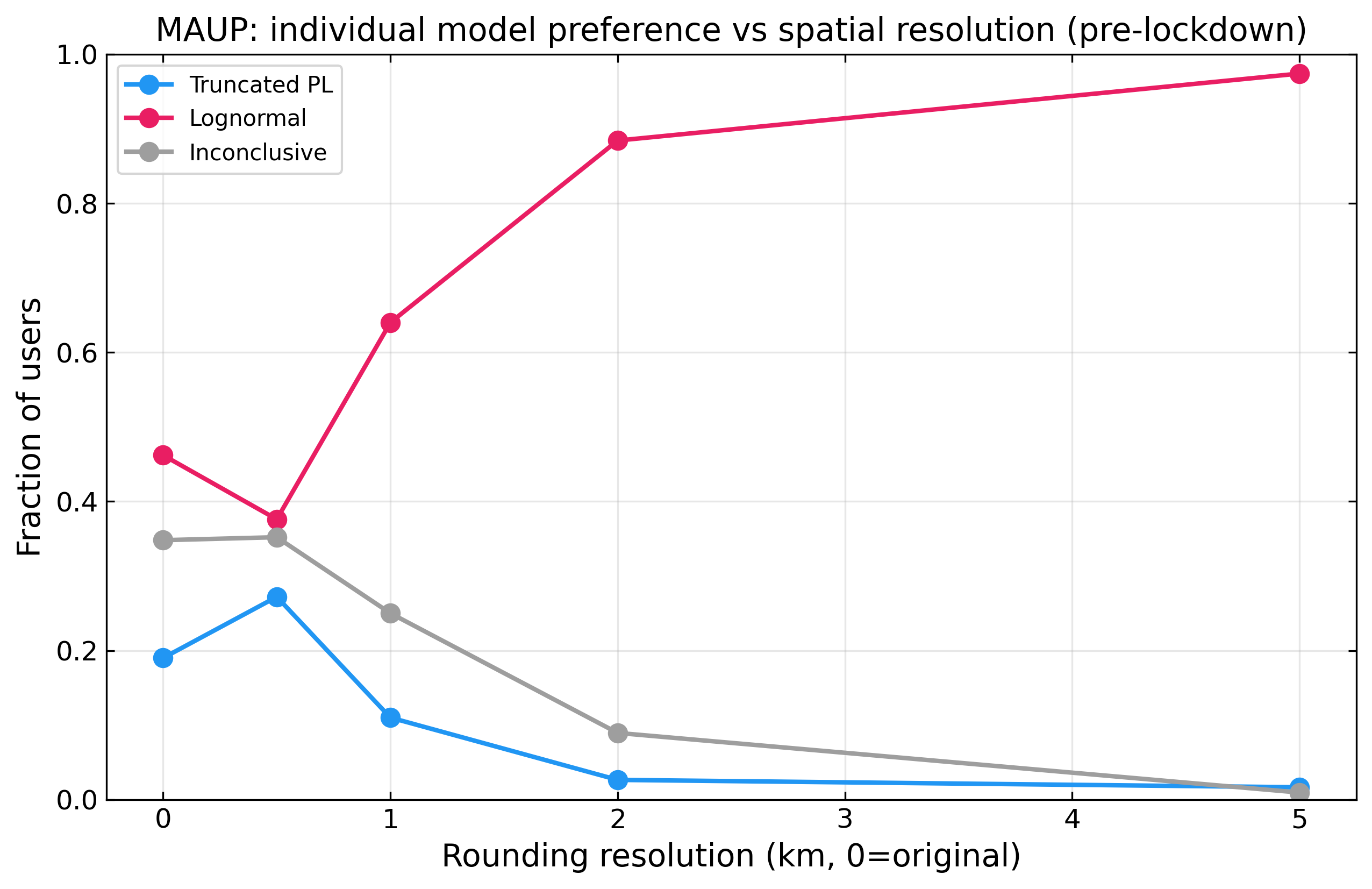}{0.78}{fig:indivres}{Individual-level winner fractions against spatial resolution. Coarser resolution strengthens the lognormal preference.}

\section{Within-level threshold sensitivity}
The main text reports that local, neighbourhood-scale displacements are
lognormal, and that intermediate scales are mixed. We test this with five
threshold sets at the aggregate level, where the tests have full power. The
shortest level (below 5, 3, or 2.5~km) favours lognormal across all sets. The
longest level (above 300~km) also favours lognormal. The intermediate levels (5
to 50~km and similar) often favour truncated power law. Across all 23
within-level fits, lognormal wins 11 and power law wins 12
(Fig.~\ref{fig:withinlevel}). The individual within-level fits in the main panel
use scarce per-user-per-level data, which moves toward lognormal or
inconclusive. The aggregate fits here pool millions of points and resolve the
mixed middle. Table~\ref{tab:ed1} reports the aggregate within-level fits for
the original threshold set.

\begin{table}[ht]\centering
\caption{Within-level model comparison (aggregate, pre-lockdown). Positive $R$
favours lognormal. $\Delta$AIC $=$ AIC(power law) $-$ AIC(lognormal).}
\label{tab:ed1}
\begin{adjustbox}{width=\linewidth}
\begin{tabular}{lrrrrr}
\toprule
Level & $N$ & \% & $R$(LN/PL) & $\Delta$AIC & $\alpha_{\mathrm{PL}}$ \\
\midrule
Neighbourhood ($<5$ km) & 340{,}985{,}722 & 73.5\% & $+100{,}070$ & $+200{,}138$ & 1.108 \\
City (5--50 km)         & 116{,}393{,}937 & 25.1\% & $+3{,}543$   & $+7{,}084$   & 2.258 \\
Region (50--300 km)     & 3{,}947{,}837   & 0.9\%  & $+713$       & $+1{,}423$   & 2.847 \\
Long-distance ($>300$ km)& 2{,}632{,}742  & 0.6\%  & $+3{,}043$   & $+6{,}083$   & 2.042 \\
\bottomrule
\end{tabular}
\end{adjustbox}
\end{table}

\sfig{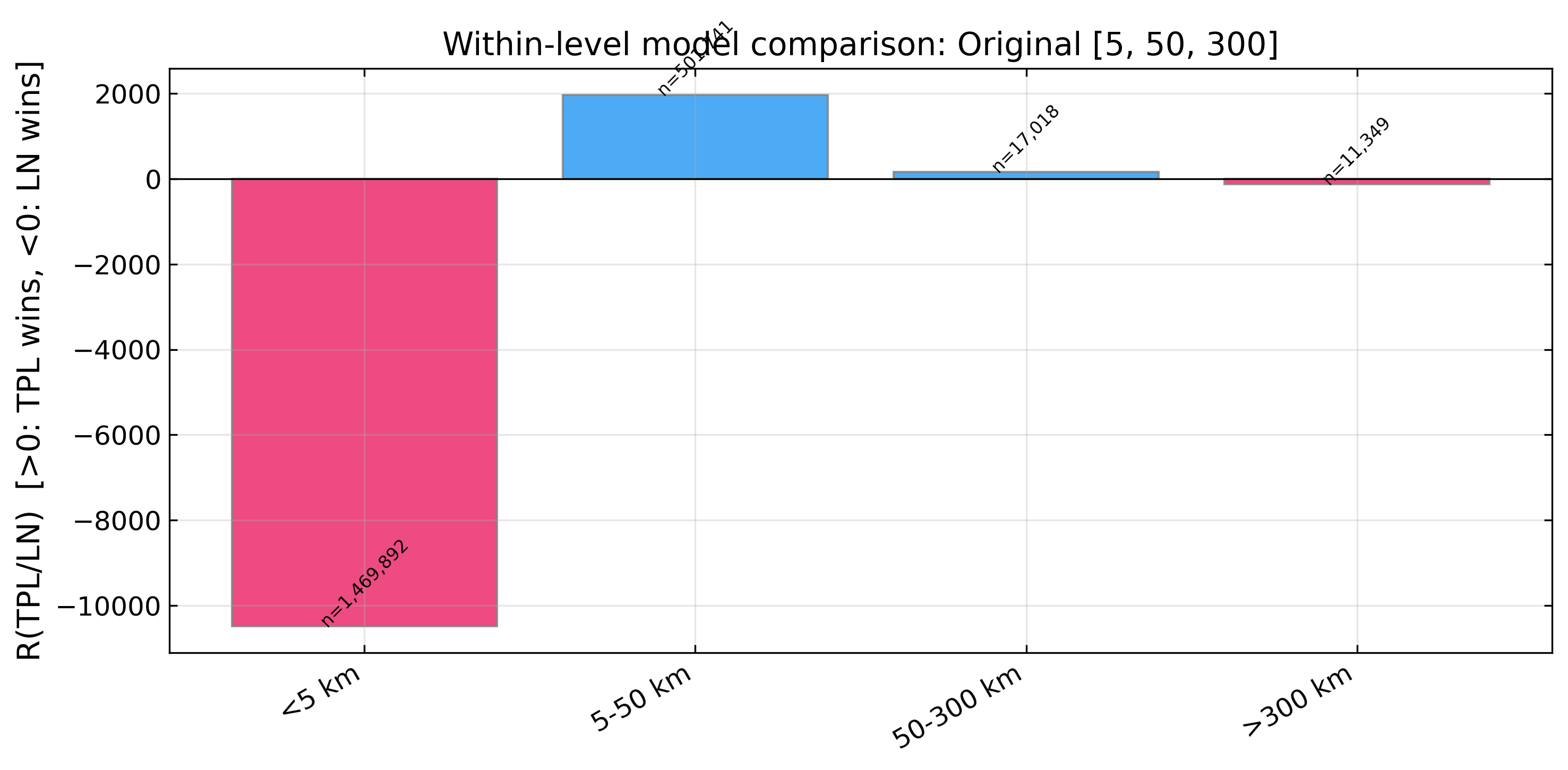}{0.5}{fig:withinlevel}{Within-level model comparison for the original
threshold set. The shortest level strongly favours lognormal, and the 5--50~km
level favours truncated power law.}

\section{Mixture decomposition}
A naive parametric mixture fits a lognormal to each hard-truncated distance bin
and sums them. This fails, with negative $R^{2}$. The lognormal density drops to
zero at bin edges and leaves gaps. The kernel density estimate removes these
gaps. The KDE level mixture reaches $R^{2}=0.977$ before lockdown and
$R^{2}=0.984$ under full lockdown (Table~\ref{tab:ed2}). The value is lower
during partial lockdown (0.615). Fixed thresholds track shifting mobility less
well in a transitional period.

\begin{table}[ht]\centering
\caption{Mixture decomposition $R^{2}$ by period. Parametric LN: a lognormal
fitted to each hard-truncated bin. KDE: kernel density estimation per level.
Free GMM: a four-component Gaussian mixture in log-space.}
\label{tab:ed2}
\begin{adjustbox}{width=\linewidth}
\begin{tabular}{lrrr}
\toprule
Period & Parametric LN & KDE level mix & Free GMM \\
\midrule
Pre-lockdown    & $-5.547$  & 0.977 & 0.898 \\
Partial lockdown& $-13.034$ & 0.615 & 0.864 \\
Full lockdown   & $-13.895$ & 0.984 & 0.863 \\
\bottomrule
\end{tabular}
\end{adjustbox}
\end{table}

\section{Generative mechanism}
A variable is lognormal when its logarithm is normal. Products of many positive
factors give a lognormal by the central limit theorem\cite{Limpert_2001}. We propose that trip
distance within a container is such a product. Destination type, routing, and
availability each act on the distance in proportion. This mechanism matches
Gibrat's law for city sizes\cite{Eeckhout_2004}. A minimal four-level container model, with each
level generating lognormal displacements, produces a power-law-like aggregate
with exponent near 1.67 (Fig.~\ref{fig:sim}). Formal normality tests on log
displacements reject exact normality at every level
(Fig.~\ref{fig:qq}). The within-level law is close to lognormal, not exactly
lognormal. We present the mechanism as an approximation.

\sfig{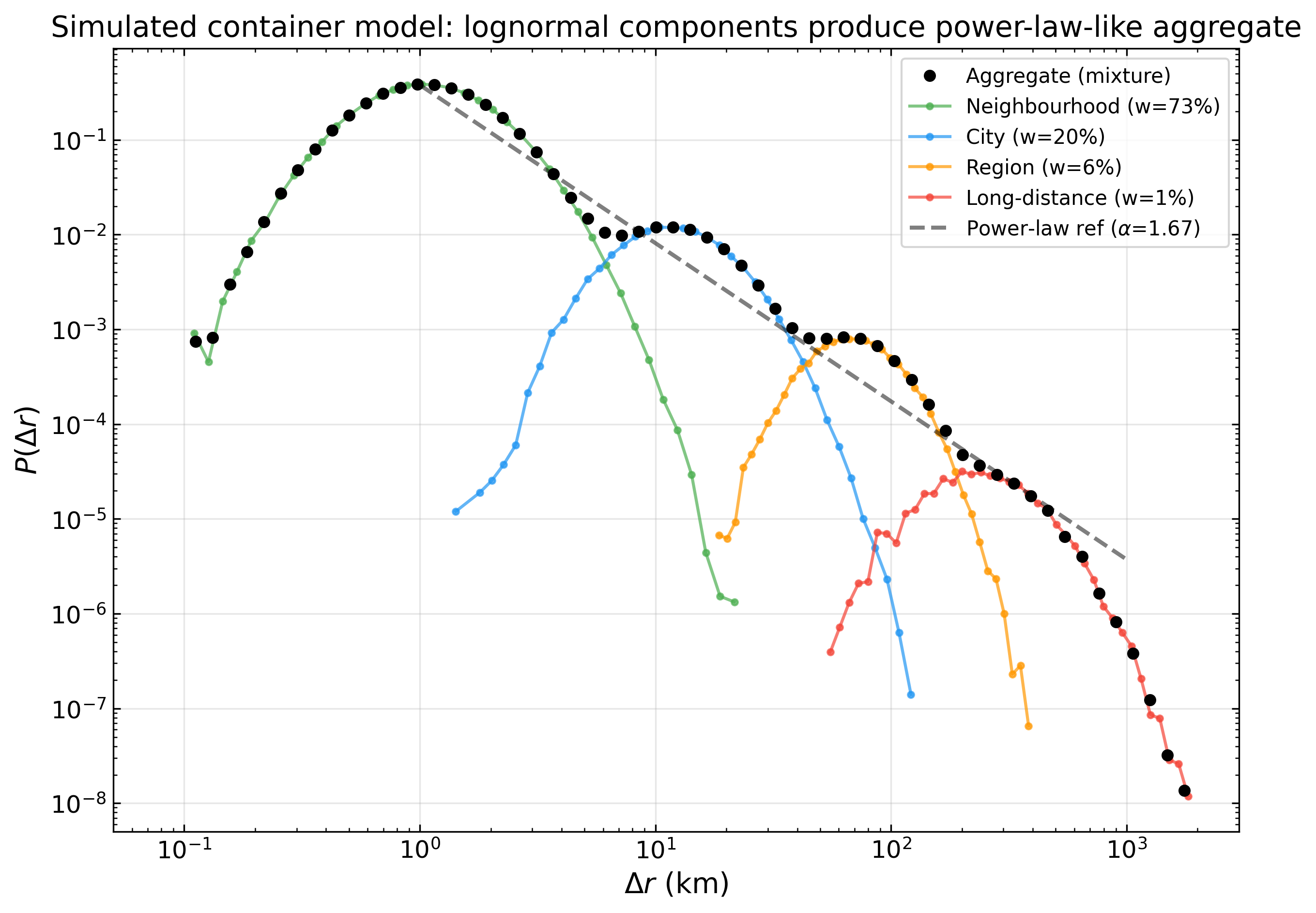}{0.5}{fig:sim}{Simulated container model. Lognormal components
(coloured curves) sum to a power-law-like aggregate (black points). The dashed
line shows a power-law reference with $\alpha=1.67$.}

\sfig{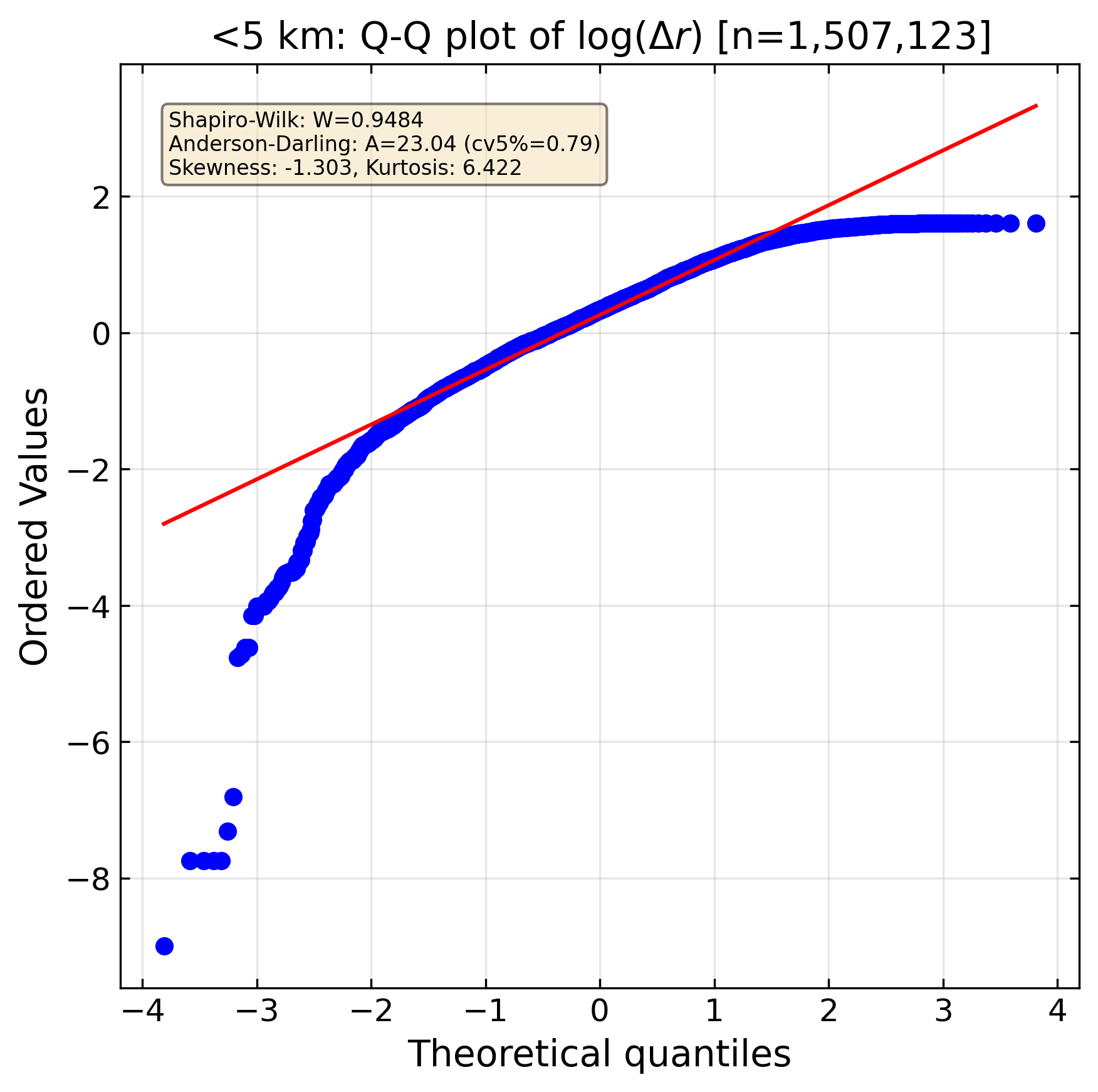}{0.5}{fig:qq}{Quantile--quantile normal plot for log displacements at
the $<5$~km level ($n=1.5$M). The body is near-normal. Both tails deviate.}

\section{Inter-event times and return probabilities}
We sampled 500{,}000 inter-event times per period. The median falls modestly
from 1{,}348~s before lockdown to 1{,}227~s under full lockdown. The
inter-event exponent is 0.41, lower than the 0.9 of Gonz\'alez et al.
\cite{Gonzalez_2008}. Our data are XDR (all network events), not CDR (calls and
texts), which gives more frequent events. This is a data-type difference, not a
mobility difference. We also computed return probabilities for 5{,}000 users
before lockdown. The return probability rises to about 0.90 by step 100
(Fig.~\ref{fig:returnprob}), close to the value reported by Gonz\'alez et al.
Tower identifiers exist only for the pre-lockdown export, so we cannot compare
return probabilities across periods.

\sfig{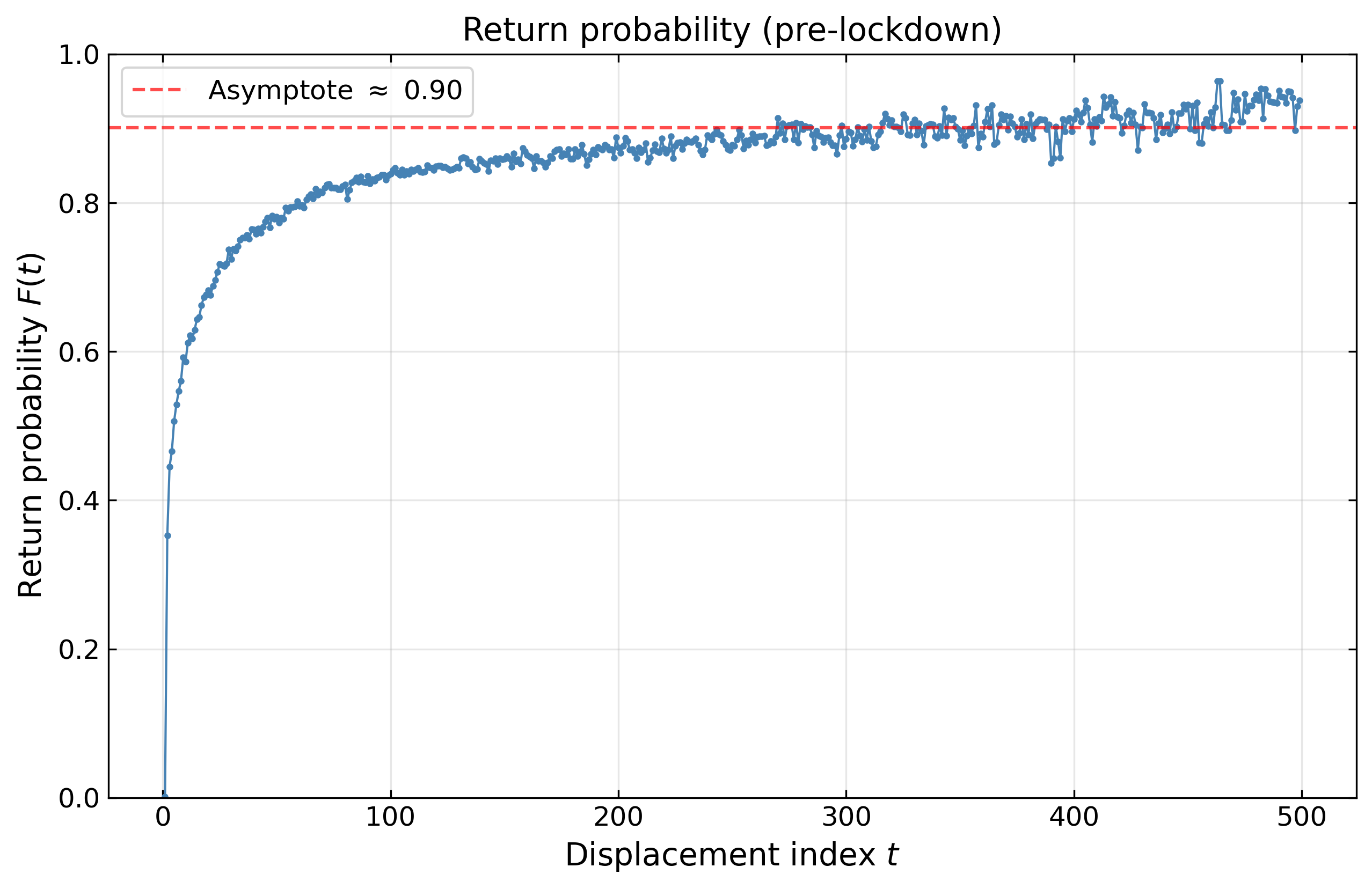}{0.5}{fig:returnprob}{Return probability $F(t)$ for the pre-lockdown
period (5{,}000 users). The plateau near 0.90 marks highly recurrent mobility.}

\section{Replication of Gonz\'alez et al. (2008)}
We replicated the Gonz\'alez et al.\ (2008) supplementary analyses on the
Chilean data (Table~\ref{tab:gonzalez}). The spatial results match well. The
displacement exponent is 1.74, against 1.75. Trajectory universality and
sub-diffusive $r_{g}$ growth both reproduce. The temporal results differ, in line
with XDR versus CDR sampling. The standard KS test rejects the truncated power
law for both $P(\Delta r)$ and $P(r_{g})$. The weighted KS test, more sensitive in the
tails, does not reject. This split fits our thesis. The aggregate distribution
departs from a power law in the body, where lognormal parts curve it, and
matches in the tails.

\begin{table}[ht]\centering
\caption{Replication of Gonz\'alez et al.\ (2008) supplementary analyses.}
\label{tab:gonzalez}
\begin{adjustbox}{width=\linewidth}
\begin{tabular}{cllll}
\toprule
Fig & Analysis & Gonz\'alez & Our data & Notes \\
\midrule
S1 & Interevent exponent $\alpha$ & 0.9 & 0.41 & XDR vs CDR \\
S2 & $P(\Delta r\mid\Delta T_0)$ exponent & 1.75 & 1.67 & within error \\
S5 & KS test $P(\Delta r)$  & $p=1.00$ & $p=0.000$ & TPL rejected \\
S6 & KSW test $P(\Delta r)$ & $p=1.00$ & $p=0.148$ & borderline \\
S7 & KS test $P(r_{g})$  & $p=0.62$ & $p=0.000$ & TPL rejected \\
S8 & KSW test $P(r_{g})$ & $p=0.82$ & $p=0.729$ & consistent \\
S9 & $P(\Delta r)$ exponent $\beta$ & 1.75 & 1.74 & exact replication \\
\bottomrule
\end{tabular}
\end{adjustbox}
\end{table}

\section{Tail-test machinery and calibration}
All tail tests reduce the power-law question to a test of exponentiality. A
variable that is Pareto above a threshold has an exponential log-excess. A
lognormal variable has a truncated-normal log-excess. The coefficient of
variation of an exponential sample equals 1 at every threshold inside a true
Pareto tail. Departures from 1 are departures from Pareto. We use two tests.
The first is the uniformly most powerful unbiased test of del Castillo and Puig
\cite{del_Castillo_1999}, which detects under-dispersion. The second is the
maximum-entropy likelihood-ratio test of Bee and colleagues \cite{Bee_2011}. We
also trace the raw coefficient of variation with a two-sided statistic, which
responds to both over- and under-dispersion. We checked the machinery on
synthetic data. On true Pareto tails the test rejects at the nominal 5\% rate.
On a lognormal sample the top 100 points pass as Pareto in 90\% of seeds, and
the top 500 reject in half. This is the small-sample illusion \cite{Perline_2005}.

\section{The extreme tail is a set of corridors}
The largest displacements are a few repeated tower-pair distances, not a
continuum. The concentration is exact at full floating-point precision
(Table~\ref{tab:corridor}). Under full lockdown the 100 longest displacements in
a 2-million sample take only 7 distinct values. The single most common value,
2{,}418.3~km, occurs 40 times. The number of distinct corridors in the top 100
falls from 67 before lockdown to 7 under it. These corridor atoms (above about
1{,}000~km, 0.3\% of records) sit outside any continuous tail model. We drop
them from the continuous tests and treat them on their own here.

\begin{table}[ht]\centering
\caption{Concentration of the extreme tail on travel corridors, by period.}
\label{tab:corridor}
\begin{adjustbox}{width=\linewidth}
\begin{tabular}{lrrrr}
\toprule
Period & top-100 distinct & top-10 share (top-100) & top-1000 distinct & top-10 share (top-1000) \\
\midrule
Pre     & 67 & 41\%  & 261 & 37\% \\
Partial & 7  & 100\% & 167 & 49\% \\
Full    & 7  & 100\% & 140 & 50\% \\
\bottomrule
\end{tabular}
\end{adjustbox}
\end{table}

\section{Threshold scan at full power}
We scan the tail threshold rank by rank at the full 2-million sample size. The
one-sided test rejects the power law for thresholds from about 39 to 1{,}020~km,
with $p$ below $10^{-40}$. Near 1{,}330 to 1{,}580~km the scan sits inside the
corridor atoms, where continuous tests do not apply. Below about 35~km the
log-excess is over-dispersed, and the one-sided test returns $p\approx 1$ by
construction. The two-sided test rejects the power law there at $p<10^{-11}$
(Fig.~\ref{fig:twotests}). Read together, the two tests show no Pareto regime at
any threshold.

\sfig{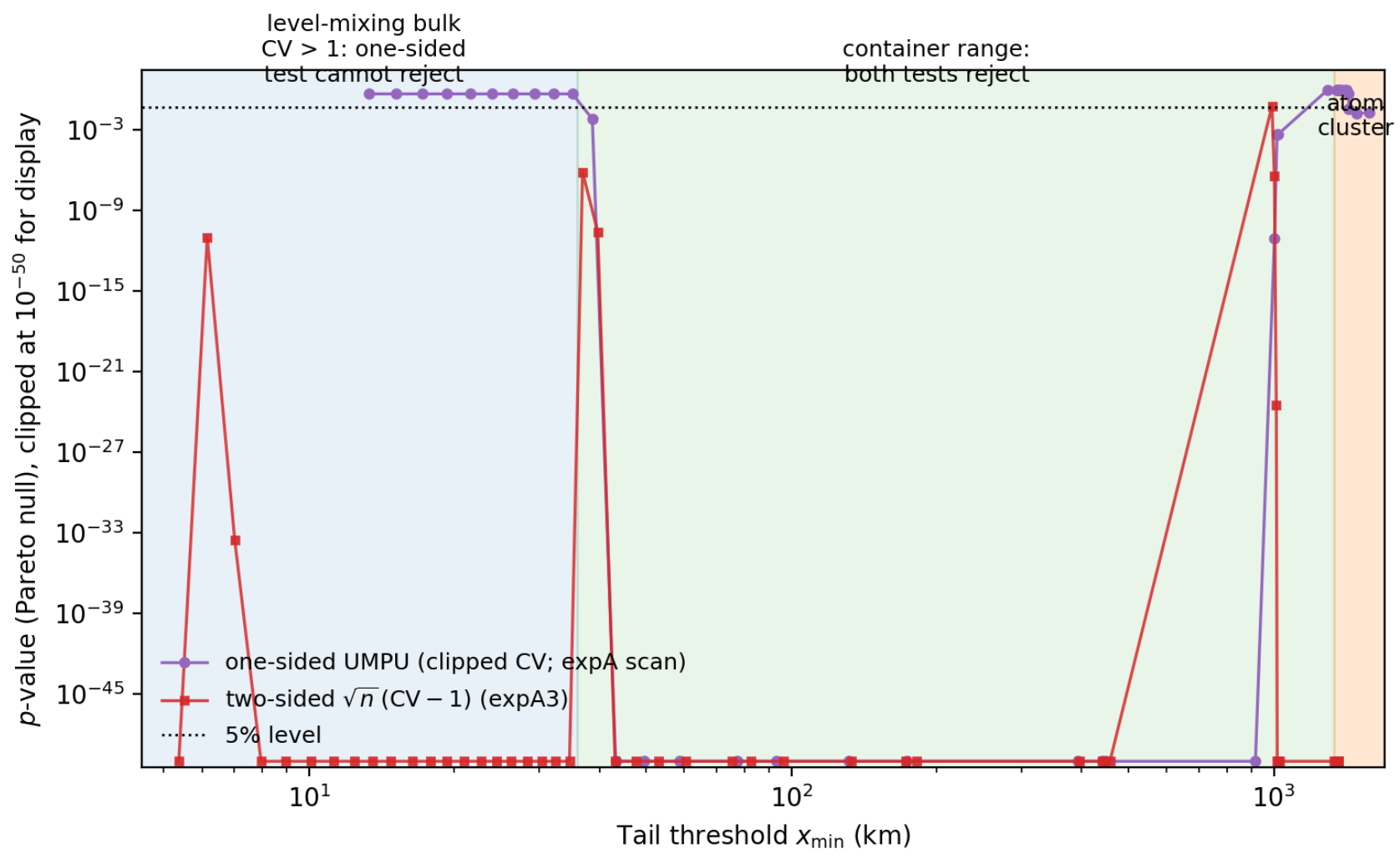}{0.85}{fig:twotests}{The two tests across the threshold (pre-lockdown,
$p$-values clipped at $10^{-50}$ for display). The one-sided test (purple)
rejects in the container range. It is blind to the over-dispersed bulk below
about 35~km, where it returns $p\approx 1$. The two-sided test (red) rejects
there. The atom cluster near 1{,}356~km is where continuous tests do not apply.}

\section{The small-sample illusion}
Perline and Bee showed that the top points of a lognormal sample look Pareto at
large $N$ and not Pareto at small $N$ \cite{Perline_2005,Bee_2011}. We subsample
the pre-lockdown aggregate from $10^{3}$ to $2\times10^{6}$. The apparent Pareto
tail fraction is 14 to 16\% at a few thousand points and gone by $N=10^{4}$
(Table~\ref{tab:tailfrac}, Fig.~\ref{fig:tailfrac}). The model-comparison
verdict is stable with scale. The Vuong test prefers lognormal at every $N$,
more decisively as $\sqrt{N}$, from $R=-3.1$ at $N=10^{3}$ to $R=-155$ at
$N=2\times10^{6}$ (Fig.~\ref{fig:vuongN}). The fitted exponent stays near 1.66
throughout. The classic studies fitted samples of order $10^{5}$, the range
where a lognormal mixture looks Pareto.

\begin{table}[ht]\centering
\caption{Apparent Pareto-compatible tail fraction $k^{*}/N$ against sample size,
pre-lockdown.}
\label{tab:tailfrac}
\begin{adjustbox}{width=\linewidth}
\begin{tabular}{lrrrrr}
\toprule
$N$ & $10^{3}$ & $3\times10^{3}$ & $10^{4}$ & $3\times10^{4}$ & $\geq 10^{5}$ \\
\midrule
$k^{*}/N$ & 0.142 & 0.159 & 0 & 0.002 & 0 \\
\bottomrule
\end{tabular}
\end{adjustbox}
\end{table}

\sfig{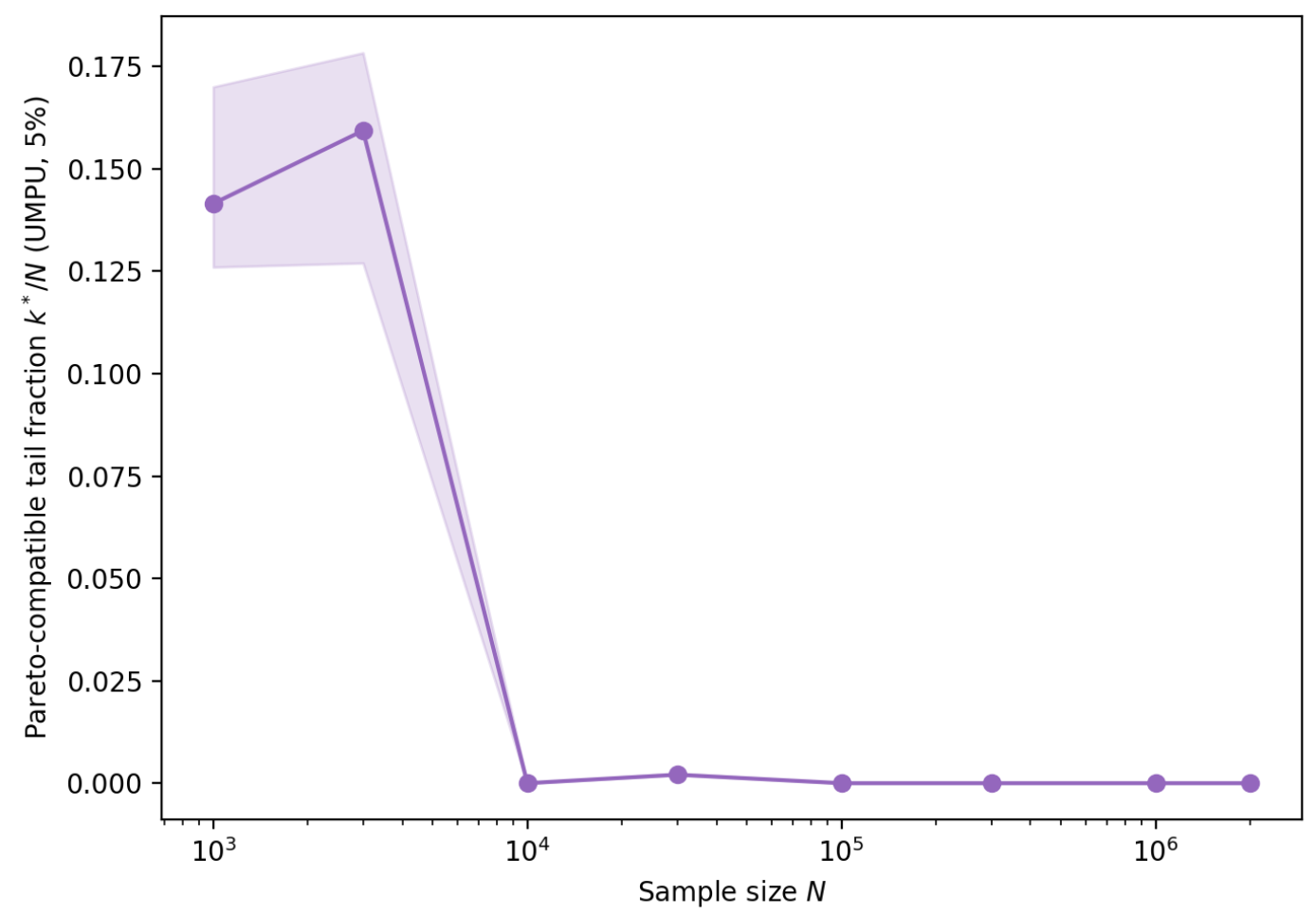}{0.5}{fig:tailfrac}{The Pareto-compatible tail fraction reported by the
threshold scan, against sample size. A few-thousand-point sample appears to
carry a Pareto tail over 14 to 16\% of its range. By $N=10^{4}$ it has vanished.}

\sfig{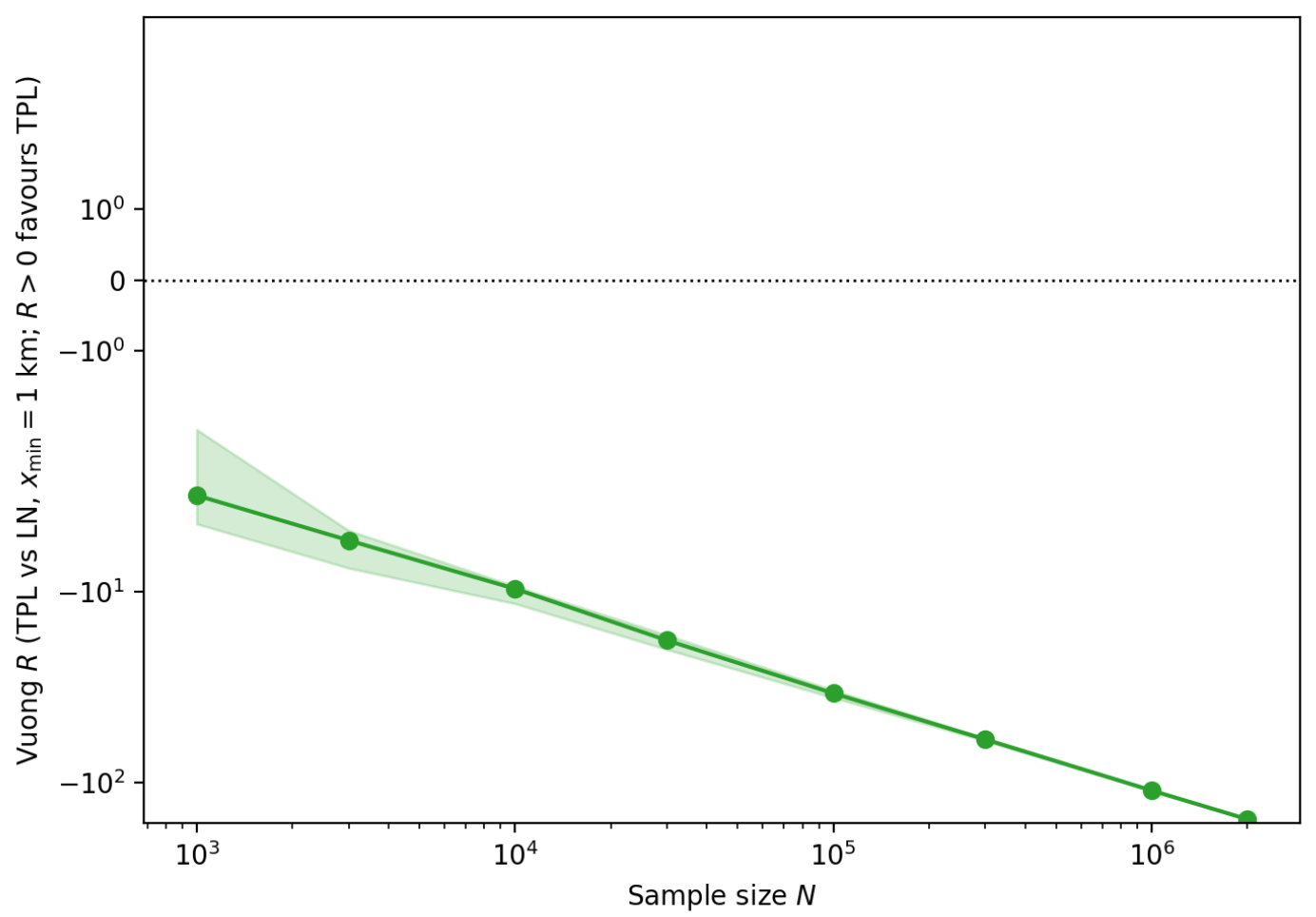}{0.5}{fig:vuongN}{The Vuong model-comparison statistic (truncated power
law vs lognormal, $x_{\min}=1$~km) against sample size. Lognormal is preferred
at every $N$, more decisively as $\sqrt{N}$. The classification verdict is
scale-stable.}

\section{Matched-size aggregation}
We compare a single user's tail with a pooled population sample of the same
size. For 1{,}537 users with at least 300 pre-lockdown displacements, we test
the top-20\% tail (Table~\ref{tab:matched}). Individual tails reject Pareto
52.9\% of the time. Pooled samples of matched size reject only 16.7\% of the
time. The pooled Pareto tail runs about three times longer. Two checks hold the
result in place. Repeated commute distances do not drive the rejections. The
user rejection rate is flat across duplicate-share quartiles, and the 55 users
with almost no duplicates still reject at 51\% against 13\% for their pooled
matches. Tail censoring makes the three-times figure a lower bound. 45\% of
pooled samples hit the length cap, against 22\% of users.

\begin{table}[ht]\centering
\caption{Matched-size comparison of the top-20\% tail, individual users versus
pooled samples of the same size.}
\label{tab:matched}
\begin{adjustbox}{width=\linewidth}
\begin{tabular}{lrrr}
\toprule
Metric (top-20\% tail) & Individual & Pooled & Wilcoxon $p$ \\
\midrule
Tail-test $p$-value (median)     & 0.027 & 0.476 & $2\times10^{-50}$ \\
Pareto rejected at 5\% (share)   & 52.9\% & 16.7\% & --- \\
Pareto tail fraction (median)    & 0.096 & 0.281 & $4\times10^{-137}$ \\
Clipped CV (median; Pareto $=1$) & 0.801 & 0.973 & $4\times10^{-106}$ \\
\bottomrule
\end{tabular}
\end{adjustbox}
\end{table}

\section{Composition reweighting and the H1 null}
We reweight pre-lockdown users so their $r_{g}$ distribution matches the
full-lockdown one. We pool only their pre-lockdown displacements and refit. The
reweighting is well-behaved, with an effective sample of 5{,}788 and a KS
distance to the target falling from 0.27 to 0.02. The fitted exponent moves from
1.673 to 1.814, a composition-only shift of $+0.141$. This overshoots the
observed panel shift of $+0.065$. We then apply the same reweighting to
simulated H1 worlds built from the panel's real $r_{g}$ values and displacement
counts. A pure, untruncated power law is invariant, which confirms the
machinery. Every realistic truncated-L\'evy kernel shifts under reweighting, by
$+0.04$ to $+0.18$ (Table~\ref{tab:h1null}, Fig.~\ref{fig:h1shift}). These
bracket the observed value. The composition result is a sufficiency statement,
not a differential test. The aggregate exponent shift cannot by itself separate
the two accounts.

\begin{table}[ht]\centering
\caption{Exponent shift under the same reweighting, for simulated H1 worlds with
a single universal exponent $\beta=1.66$.}
\label{tab:h1null}
\begin{adjustbox}{width=\linewidth}
\begin{tabular}{lr}
\toprule
H1 kernel & Shift in $\alpha$ \\
\midrule
Pure power law, support $\ll x_{\min}$ (theoretical check) & $-0.002\pm0.006$ \\
Truncated, cutoff $10\,r_{g}$, support $\ll x_{\min}$        & $+0.036\pm0.009$ \\
Pure power law, inner scale $0.05\,r_{g}$                    & $+0.066\pm0.001$ \\
Truncated, cutoff $10\,r_{g}$, inner scale $0.05\,r_{g}$       & $+0.145\pm0.002$ \\
Truncated, cutoff $3\,r_{g}$, inner scale $0.05\,r_{g}$        & $+0.177\pm0.001$ \\
\bottomrule
\end{tabular}
\end{adjustbox}
\end{table}

\sfig{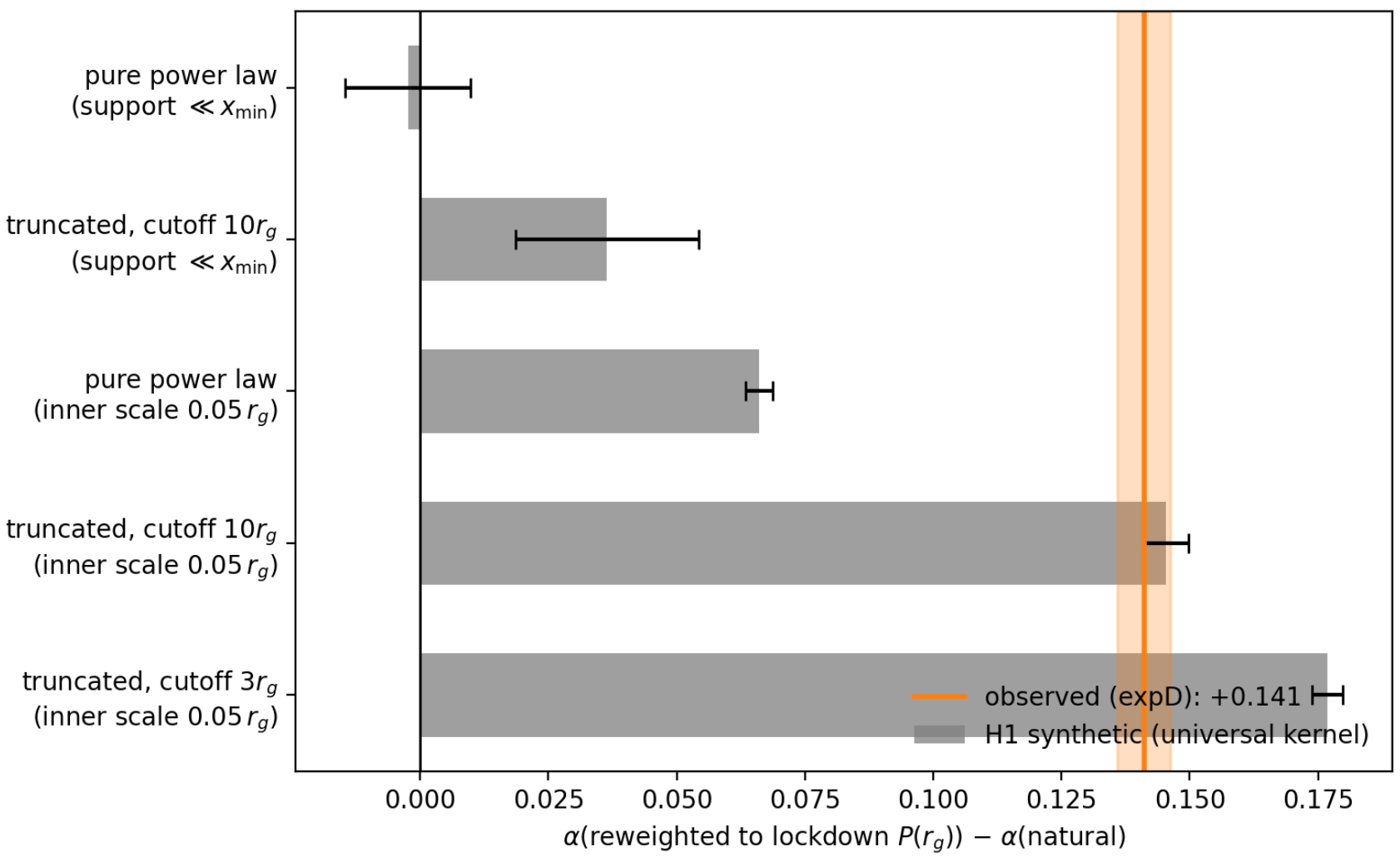}{0.5}{fig:h1shift}{H1-predicted exponent shifts under the same
reweighting (grey bars, $\pm2\sigma$), against the observed shift (orange line
and band). A pure, untruncated power law is invariant. Realistic truncated-Lévy
kernels shift by amounts that bracket the observed value. The reweighting shows
sufficiency, not a differential test.}


\end{document}